\documentclass[]{emulateapj}

\usepackage{rotating}
\usepackage{subfigure}
\usepackage{longtable}
\usepackage{url}
\usepackage{hyperref}
\usepackage{morefloats}
\usepackage{amsmath} 

\newcommand{\angstrom}{\text{\normalfont\AA}}
\newcommand{\Kepler}{{\it Kepler~}}

\newcommand{\noprint}[1]{}

\newcommand{\teff}{$T_{\rm eff}$~}

\begin{document}

\title{Kepler-445, Kepler-446 and the Occurrence of Compact Multiples Orbiting Mid-M Dwarf Stars}

\author{Philip S. Muirhead,\altaffilmark{1} Andrew W. Mann,\altaffilmark{2,10,11} Andrew Vanderburg,\altaffilmark{3} Timothy D. Morton,\altaffilmark{4} Adam Kraus,\altaffilmark{2} Michael Ireland,\altaffilmark{5} Jonathan J. Swift,\altaffilmark{6} Gregory A. Feiden,\altaffilmark{7} Eric Gaidos,\altaffilmark{8} J. Zachary Gazak\altaffilmark{9}}

\affil{\vspace{0pt}\\ $^{1}$Department of Astronomy, Boston University, 725 Commonwealth Ave., Boston, MA, 02215 USA \\
$^{2}$Department of Astronomy, The University of Texas at Austin, Austin, TX 78712, USA \\
$^{3}$Harvard-Smithsonian Center for Astrophysics, 60 Garden St., Cambridge, MA 02138,  USA \\
$^{4}$Department of Astrophysical Sciences, 4 Ivy Lane, Peyton Hall, Princeton University, Princeton, NJ 08544, USA \\
$^{5}$Australian National University, Canberra ACT 2611, Australia \\
$^{6}$Department of Astrophysics, California Institute of Technology, MC 249-17, Pasadena, CA 91125, USA \\
$^{7}$Department of Physics and Astronomy, Uppsala University, Box 516, Uppsala 751 20, Sweden \\
$^{8}$Department of Geology and Geophysics, University of Hawai'i at Manoa, Honolulu, HI 96822, USA \\
$^{9}$Institute for Astronomy, University of Hawai'i at Manoa, 2680 Woodlawn Drive, Honolulu, HI 96822, USA}

\altaffiltext{10}{Harlan J. Smith Fellow, The University of Texas at Austin}
\altaffiltext{11}{Visiting Researcher, Institute for Astrophysical Research, Boston University}

\begin{abstract}

We confirm and characterize the exoplanetary systems {\it Kepler}-445 and {\it Kepler}-446: two mid-M dwarf stars, each with multiple, small, short-period transiting planets.  {\it Kepler}-445 is a metal-rich ([Fe/H]=+0.25 $\pm$ 0.10) M4 dwarf with three transiting planets, and {\it Kepler}-446 is a metal-poor ([Fe/H]=-0.30 $\pm$ 0.10) M4 dwarf also with three transiting planets.  {\it Kepler}-445c is similar to GJ 1214b: both in planetary radius and the properties of the host star.  The {\it Kepler}-446 system is similar to the {\it Kepler}-42 system: both are metal-poor with large galactic space velocities and three short-period, likely-rocky transiting planets that were initially assigned erroneously large planet-to-star radius ratios.  We independently determined stellar parameters from spectroscopy and searched for and fitted the transit light curves for the planets, imposing a strict prior on stellar density in order to remove correlations between the fitted impact parameter and planet-to-star radius ratio for short-duration transits.  Combining {\it Kepler}-445, {\it Kepler}-446 and {\it Kepler}-42, and isolating all mid-M dwarf stars observed by {\it Kepler} with the precision necessary to detect similar systems, we calculate that 21 $^{+7}_{-5}$ \% of mid-M dwarf stars host compact multiples (multiple planets with periods of less than 10 days) for a wide range of metallicities.  We suggest that the inferred planet masses for these systems support highly efficient accretion of protoplanetary disk metals by mid-M dwarf protoplanets.

\end{abstract}

\keywords{stars: fundamental parameters --- stars: individual ({\it Kepler}-445, {\it Kepler}-446, {\it Kepler}-42, Barnard's Star) --- stars: late-type --- stars: low-mass -- stars: planetary systems}

\maketitle

\section{Introduction}

The {\it Kepler}-42 exoplanetary system is rather remarkable: it consists of three sub-Earth-sized planets all orbiting and transiting a mid-M dwarf host star with periods of less than two days \citep{Muirhead2012a}, a so-called ``compact multiple.''  Consider that {\it Kepler}-42 is a metal-poor star, with a measured [M/H] of -0.27 in \citet{Muirhead2012a}.  Following the calculation of \citet{Schlaufman2010}, if we assume the planets formed from material in a protoplanetary disk with a total disk mass equal to 1\% of the current host star mass, and that the disk had identical metal abundance to the star today, that leads to a disk metal content totaling 4.1 Earth masses.  Assuming the three planets have primarily rocky compositions, and combining the predicted planetary mass-radius relationships of \citet{Fortney2007} with the measured planetary radii in \citet{Muirhead2012a}, the total mass of the three planets is calculated to be 0.71 Earth masses.  Under these admittedly simplistic assumptions, nearly 20\% of {\it Kepler}-42's disk metals went into the formation of these three rocky planets.  The same calculation for the Sun results in only 5\% of disk metals contributing to rocky planets (Earth, Venus, Mars and Mercury), with significantly more contributing to the cores of the solar system's gas giant planets.  

The preference for metals to contribute to rocky planets rather than gas-giant cores would be strong evidence for the planet-formation scenario suggested by \citet{Laughlin2004}, in which gas-giant-core embryos form in the protoplanetary disks around M dwarf stars; however, the gas in the disk dissipates before those embryos grow large enough to accrete and are cut-off as terrestrial planets.  The scenario is already supported by the relative scarcity of gas-giant exoplanets found to orbit M dwarf stars.  Using radial velocity observations, \citet{Johnson2010} found a statistical decrease in giant planet planet occurrence with decreasing host star mass, including M dwarfs in the radial velocity sample.  However, \citet{Gaidos2014b} do not find strong support for a statistical deficiency of gas giant planets orbiting M dwarfs, though they cannot statistically rule out a deficiency.  Regardless, the presence of failed embryos in some consistent proportion to the amount of available metals in the protoplanetary disk would provide support for the cut-off accretion scenario.

It is important to determine whether the planetary system orbiting {\it Kepler}-42 is an outlier or a common type of system around mid-M dwarf stars, and whether metallicity affects the sizes of failed embryos.  The occurrence of compact multiple systems orbiting mid-M dwarf stars has not been thoroughly analyzed.  Recent investigations aimed at understanding the planet occurrence statistics from {\it Kepler}, such as \citet{Howard2012}, \citet{Petigura2013} and \cite{Silburt2014} do not consider mid-M dwarf stars. \citet{Dressing2013}, \citet{Gaidos2014} and \citet{Morton2014} investigated the occurrence of planets around specifically M dwarf stars and found a propensity for Earth-sized planets, the latter finding roughly two planets per M dwarf star with periods of less than 150 days.  However, those investigations did not address the occurrence of compact multiples, and many {\it Kepler} mid-M dwarf stars were not included in that study, including {\it Kepler}-42.  \citet{Swift2013} investigated the occurrence of compact multiples orbiting \Kepler M dwarf stars using {\it Kepler}-32 as a representative of the \Kepler M dwarf stars.  They found that \Kepler planet detections around M dwarf stars could be recreated reasonably well assuming all planetary systems are clones of the {\it Kepler}-32 system, indicating a high occurrence of compact multiples.

In this paper we confirm and characterize two new compact multiple systems with mid-M-dwarf host stars: {\it Kepler}-445\footnote{KOI-2704, KIC 9730163, $\alpha=298.736115^\circ$, $\delta=46.498634^\circ$} and {\it Kepler}-446\footnote{KOI-2842, KIC 8733898, $\alpha=282.250214^\circ$, $\delta=44.921108^\circ$}, initially discovered by the {\it Kepler} exoplanet search pipeline to host two and three short-period planets, respectively.  {\it Kepler}-445c, {\it Kepler}-445b, {\it Kepler}-446b and {\it Kepler}-446d were first reported as planet-candidates by \citet{Burke2014} as KOI-2704.01, KOI-2704.02, KOI-2842.01 and KOI-2842.02 respectively, and {\it Kepler}-446c was reported as a threshold-crossing event by \citet{Tenenbaum2013}, and later added to the NASA Exoplanet Archive as KOI-2842.03.   With regard to the host stars, {\it Kepler}-445 was identified as a mid-M dwarf star in an optical spectroscopic survey of late-type KOIs by \citet{Mann2013c}, and both {\it Kepler}-445 and {\it Kepler}-446 were identified as mid-M dwarf stars in a separate infrared spectroscopic survey of late-type KOIs by \citet{Muirhead2014}.  We searched the light curves for additional planets, and found an additional planet orbiting {\it Kepler}-445 not detected by the {\it Kepler} pipeline, which we refer to as {\it Kepler}-445b.  Refining the parameters for the host stars and refitting the planet transit light curves, we find that {\it Kepler}-445c is similar to GJ 1214b: both are likely mini-Neptunes orbiting metal-rich mid-M dwarf stars, and that {\it Kepler}-446 is similar to {\it Kepler}-42 in multiple ways: the low-metallicity of the host star, the multiplicity, sizes, and orbital periods of the orbiting planets, and the planets' initial mischaracterization.

We present the observations, data and analysis of these systems in Section~\ref{sec:data}. In Section \ref{sec:fpp} we calculate the false-positive probabilities for the planets orbiting {\it Kepler}-445 and {\it Kepler}-446, confirming the planetary nature of the transits.  In Section \ref{occurrence} we combine {\it Kepler}-445, {\it Kepler}-446 and {\it Kepler}-42 with the full sample of mid-M dwarf stars observed by {\it Kepler} with similar precision from quarters 1 to 16 to estimate the occurrence of compact multiple systems around mid-M dwarf stars, which we define as two or more planets oribiting with periods of less than 10 days.  Finally, in Section \ref{discussion}, we discuss the implications of these systems for planet formation scenarios and accretion of disk metals by protoplanets orbiting mid-M dwarf stars.

\section{Data and Analysis}\label{sec:data}

\subsection{High-contrast Imaging}\label{sec:ao}

High-contrast imaging of {\it Kepler} planet-candidates serves two purposes: with high-contrast images, the level of contamination by unresolved objects within the {\it Kepler} aperture can be determined, and in the event that there are no contaminating objects within the aperture, high-contrast imaging provides constraints on false-positive scenarios that could mimic the detected transit signal \citep[e.g.][]{Adams2012, Lillo-Box2012, Law2013, Lillo-Box2014, Dressing2014}.

We observed {\it Kepler}-445 and {\it Kepler}-446 with the Keck II telescope at Mauna Kea Observatories using the facility near-infrared adaptive-optics imager NIRC2, operated in $K'$ band.  We observed the stars using both conventional imaging and using non-redundant aperture masks, which can be placed near an image of the telescope pupil within the NIRC2 cryostat.  Non-redundant aperture masks enable high-contrast imaging by measuring individual spatial frequencies corresponding to each baseline produced by a pair of apertures \citep[][]{Haniff1987,Nakajima1989,Tuthill1999,Monnier2004,Lloyd2006,Martinache2009,Kraus2012,Ireland2013}.  By making the apertures non-redundant, high fringe-contrast and phase information is not lost to incoherent addition of fringes from redundant baselines, enabling contrast performance better than the full-dish diffraction limit of the telescope.  Phases from combinations of three individual baselines are combined into closure-phases, which probe azimuthal asymmetries in the image while remaining robust against phase-errors from instrument distortions or uncorrected atmospheric fluctuations.

We observed {\it Kepler}-445 on UT 17 July 2013 and {\it Kepler}-446 on UT 29 July 2014.  For each target, we acquired three, conventional $K'$ band AO images, each with 20-second integrations, each dithered across the detector, using multiple-correlated sampling with a Fowler depth of 16 \citep{Fowler1990}.  The non-redundant aperture masking images were acquired in the same manner as \citet{Kraus2012}: we used a nine-hole mask, with 1.5-meter-equivalent apertures creating baselines with separations of 1.5 to 9.2 meters.  We acquired six aperture-masked images in K band, each with 20-second integrations and using mutliple-correlated sampling.  

We reduced the data using the methods described in \citet{Kraus2008}, as well as a kernel-phase technique with the POISE calibration algorithm \citep{Ireland2013}, that included subsampling of the Keck pupil sub-apertures. The kernel-phase technique yielded marginally superior contrast limits, and those limits are reported here.

\begin{figure}[]
\begin{center}
\includegraphics[width=3.5in]{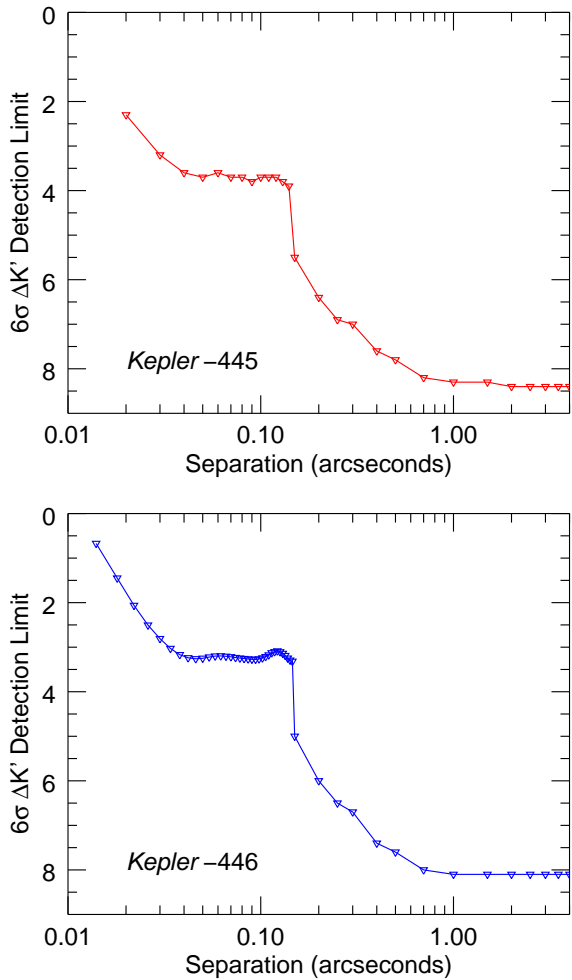}
\caption{Contrast curves for {\it Kepler}-445 ({\it top}) and {\it Kepler}-446 ({\it bottom}) showing the 6$\sigma$ $\Delta K'$ detection limits for objects at the plotted separations.  The breaks at 0.150 arcseconds arise from a switch from contrast measured using aperture-masking to that measured using conventional adaptive-optics imaging and PSF-subtraction.  We detected a faint object roughly 4 arcseconds from {\it Kepler}-445, with $\Delta K$ = 7.60 $\pm$ 0.07 magnitudes.  Such a faint object could not account for the transits seen around {\it Kepler}-445.\label{contrast_figure}}
\end{center}
\end{figure}

For {\it Kepler}-445, we detected a faint object at a separation of 3.908 $\pm$ 0.004 arcseconds, with a position angle of 279.62 $\pm$ 0.05 degrees east of north, and a magnitude difference of $\Delta K`$ = 7.60 $\pm$ 0.07 magnitudes.  With such a large contrast, the object's contribution to the {\it Kepler} light curve is negligible and it is unlikely the object showing the transit signals.  Nevertheless, we consider the possibility in our false positive analysis in Section \ref{sec:fpp}.  Aside from this detected object, we detect no other objects near {\it Kepler}-445, with 6$\sigma$ contrast limits shown in Figure~\ref{contrast_figure}.  Interestingly, this is inconsistent with the \Kepler Input Catalog \citep[KIC,][]{Batalha2010,Brown2011} and the NOMAD Catalog \citep{Zacharias2005}, both of which record a similar visible-brightness star $\sim$2."2 away: KIC 9730159 and NOMAD 1364-0341824, respectively.  Considering the infrared adaptive optics imaging, and the seeing-limited visible-wavelength imaging we report in the following section which also lacks this nearby object, we believe that the NOMAD catalog and the KIC may have inadvertently recorded {\it Kepler}-445 twice.

\subsection{Host-star Magnitudes}\label{dct}

Both {\it Kepler}-445 and {\it Kepler}-446 have reliable infrared $J$, $H$ and $Ks$ magnitudes in the Two-Micron All Sky Survey \citep[2MASS,][]{Cutri2003,Skrutskie2006}.  They are also included in the KIC, which aimed to measure SDSS-like $u$, $g$, $r$, $i$, and $z$ magnitudes for all observable stars in the {\it Kepler} field.  However, owing to their faintness, the KIC lacks $u$ and $z$ magnitudes for {\it Kepler}-445 and $u$ for {\it Kepler}-446.  Other photometric surveys of the {\it Kepler} field include the UBV Photometric Survey \citep{Everett2012}, which contains both {\it Kepler}-445 and {\it Kepler}-446 and their $B$ and $V$ magnitudes.  However, comparing the independent $V$ band measurements of {\it Kepler}-42 in \citet{Muirhead2012a} with the measurements in the UBV Survey shows a significant offset of 0.23 magnitudes, a 4$\sigma$ discrepancy, which could be due to inaccurate calibration effects for very red low-mass stars.  Yet another survey of the {\it Kepler} field is the {\it Kepler}-INT Survey \citep[KIS,][]{Greiss2012}, which contains $u$, $g$ and $r$ magnitudes for {\it Kepler}-446, but no photometry for {\it Kepler}-445.  Interestingly, \citet[][]{Greiss2012} found  a systematic offset of $\sim$0.05 magnitudes between the KIS $i$ magnitudes and the magnitudes reported in the KIC, and \citet{Pinsonneault:2012} found a similar offset between SDSS and the KIC.  We also note that the  $r$'-band Carlsberg Meridian Catalogue overlaps significantly with the {\it Kepler} field, and contains a measurement for {\it Kepler}-446 \citep[$r$' = 16.731,][]{CMC14}.   However, it does not contain a measurement for {\it Kepler}-445.

Owing to the discrepancies and incompleteness of the various surveys, we chose to measure $u$, $g$, $r$, $i$ and $z$ magnitudes of {\it Kepler}-445 and {\it Kepler}-446 independently using the Large Monolithic Imager (LMI) on the 4.3-meter Discovery Channel Telescope (DCT) in Happy Jack, Arizona \citep{Massey2013}.  LMI contains the full compliment of SDSS filters and a 6k x 6k full-wafer deep depletion e2V CCD, with a field of view of 0.25 x 0.25 degrees.  We observed {\it Kepler}-445 and {\it Kepler}-446 on UT 2014 August 6, along with two nearby SDSS fields centered on $\alpha$=290.60299, $\delta$=38.75381 and on $\alpha$=270.00583, $\delta$=270.00583, for calibration.  Conditions were photometric.  We observed each SDSS field at four different airmasses, spanning 1.00 to 1.30.  We observed {\it Kepler}-445 and {\it Kepler}-446 twice, at two separate airmasses each, with all observations between 1.00 and 1.30 airmasses.  We also observed {\it Kepler}-42 following the same strategy as {\it Kepler}-446 and {\it Kepler}-445.

For each pointing, we cycled through the SDSS filters in LMI, acquiring $u$, $g$, $r$, $i$ and $z$ band images using 2x2 pixel binning.  Each exposure was 20 seconds long in order to match or exceed the signal-to-noise of the archival SDSS photometry without saturating the LMI detector.  We performed standard flat-fielding and bias subtraction, using twilight flats for $u$ and $g$ bands, and dome flats for $r$, $i$ and $z$ bands.  To calibrate the LMI measured fluxes to SDSS magnitudes, we cross-matched all non-saturated point-sources in our SDSS calibration images with the SDSS DR9 catalog of sources \citep{Ahn2012}.  We extracted the flux for each detectable point source in the calibration images using a modified version of DAOPHOT \citep{Stetson1987}, and converted the object counts into ``detector'' magnitudes with a arbitrary zero point.  For those point sources with a reliable match in the SDSS DR9 catalog, we calculated the difference between our detector magnitudes and the SDSS archival ``PSF'' magnitudes, for each band, at each airmass.  We only used SDSS PSF magnitudes for objects that were listed as having a star-like point spread function, not saturated, with a magnitude of less than 22.  We also limited the calibrating SDSS sources to those listed as a star and having   $0.9<r-i<2.25$, similar to M3 to M6 dwarf stars \citep{Bochanski2007}.  Limiting calibration sources to stars of similar spectral type eliminates errors arising from different spectral shapes across the SDSS bands.  This resulted in 6 calibration stars for $u$ band, 83 in $g$ band, 249 in $r$ band, 465 in $i$ band and 361 in $z$ band for the SDSS calibration field with more overall sources.  For this field we fitted a line to the detector minus SDSS magnitude versus airmass for each band.  Finally, we applied the fitted relations to the measured detector magnitudes for {\it Kepler}-445, {\it Kepler}-446 and {\it Kepler}-42, again for each band.  We estimated the uncertainty in our measured magnitudes by combining the uncertainty due to photon noise with a systematic uncertainty based on the linear fit to airmass.  The estimated uncertainties are roughly consistent with the difference between magnitude determinations for the two pointings on {\it Kepler}-445 and {\it Kepler}-446.  We averaged the two pointings for {\it Kepler}-445 and {\it Kepler}-446 to obtain our best values, and the results are listed in Table \ref{magnitudes}.  We do not detect any stellar companions that would significantly contaminate the \Kepler light curves near either {\it Kepler}-445 or {\it Kepler}-446 in any of the images, despite the listed of a similar brightness star near to {\it Kepler}-445 in NOMAD and the KIC.

\begin{deluxetable}{lccc}
\tablewidth{0pt}
\tablecaption{Host-star Magnitudes\label{magnitudes}}
\tablehead{
\colhead{Band} &
\colhead{{\it Kepler}-42} & 
\colhead{{\it Kepler}-445} &
\colhead{{\it Kepler}-446}}
\startdata
$u$\tablenotemark{a} & 
19.922 $\pm$ 0.121 &
21.944 $\pm$ 0.493 &
20.929 $\pm$ 0.142 \\

$g$ &
17.142 $\pm$ 0.043 &
19.024 $\pm$ 0.032 &
18.258 $\pm$ 0.031 \\

$r$ & 
15.503 $\pm$ 0.022 &
17.626 $\pm$ 0.016 &
16.828 $\pm$ 0.016 \\

$i$ &
14.276 $\pm$ 0.015 &
16.024 $\pm$ 0.011 &
15.614 $\pm$ 0.011 \\

$z$ &
13.537 $\pm$ 0.023 &
15.087 $\pm$ 0.016 &
14.887 $\pm$ 0.016 \\[0.03in]

\tableline

$J$\tablenotemark{b} &
12.177 $\pm$ 0.021 &
13.542 $\pm$ 0.029 &	
13.591 $\pm$ 0.021 \\

$H$ & 
11.685 $\pm$ 0.018 &
12.929 $\pm$ 0.035 &
13.075 $\pm$ 0.026 \\

$K$ &
11.465 $\pm$ 0.018 &
12.610 $\pm$ 0.028 &
12.827 $\pm$ 0.024 \\[0.03in]

\tableline

$W_1$\tablenotemark{c} & 
11.240 $\pm$ 0.023 &
12.478 $\pm$ 0.024 &
12.707 $\pm$ 0.023 \\

$W_2$ & 
11.054 $\pm$ 0.021 &
12.353 $\pm$ 0.025 &
12.476 $\pm$ 0.023 \\

$W_3$ &
10.831 $\pm$ 0.059 &
11.252 $\pm$ 0.087 &
12.931 $\pm$ 0.415
\enddata
\tablenotetext{a}{$ugriz$ measured with DCT-LMI and reported as equivalent SDSS PSF AB magnitudes \citep{Oke1983}.}
\tablenotetext{b}{$J$, $H$, and $K$ from 2MASS \citep{Cutri2003}.}
\tablenotetext{c}{$W_1$, $W_2$ and $W_3$ from WISE \citep{Cutri2012}.}
\end{deluxetable}

With $r-i$ values of 1.68 and 1.28, and $i-z$ values of 0.86 and 0.64, {\it Kepler}-445 and {\it Kepler}-446 fall squarely in the color-color limits for M4 dwarf stars from \citet[][Table 1]{Bochanski2007}, who compared colors to spectral types of M dwarf stars in the SDSS survey.  We note this only for consistency.  In the next section, we use our photometry to flux calibrate and stitch together optical and infrared spectra, which provide a far greater handle on the properties of the host stars.  We also note that neither {\it Kepler}-42, {\it Kepler}-445 nor {\it Kepler}-446 shows a $u$-band excess, indicating low chromospheric activity.  This is further supported by the lack of H$\alpha$ in emission, described in the next section.

\subsection{Moderate-resolution Spectra and Stellar Parameters}

Neither {\it Kepler}-445 nor {\it Kepler}-446 currently has a published astrometric parallax measurement.  With astrometric parallaxes, relatively accurate mass-luminosity relations can be used to determine stellar masses based on their absolute infrared magnitudes \citep{Henry1993, Delfosse2000}.  The stellar masses can be combined with empirical or theoretical mass-radius relationships to determine stellar radius \citep[e.g.][]{Torres2010}.  Or, in some cases, stellar parameters can be determined from the transit light curve itself, either by measuring photometric signatures from asteroseismic pulsations \citep[e.g.][]{Bedding2011,Huber2013}, or by using an exoplanet transit light curve to infer the host star's density assuming a low-eccentricity or circular orbit \citep[e.g.][]{Seager2003, Carter2011}, or a combination of both techniques \citep[e.g.][]{Ballard2014}.

In the case of {\it Kepler}-445 and {\it Kepler}-446, asteroseismic signatures are very difficult to measure because, being M dwarf stars, they are dense, with mean densities over 10 gm/c, where the asteroseismic signals are low and the stellar oscillation frequencies are high.  Also, the signal-to-noise of the {\it Kepler} light curves is simply not high enough to provide strong constraints on the stellar density from transit light curve fitting.  Without astrometric parallaxes, asteroseismic or transit light curve constraints, we must rely on colors and spectroscopy to determine the stellar parameters.  In the case of M dwarf stars burning hydrogen on the main-sequence, spectroscopy probes primarily effective temperature, \teff, and metallicity, [Fe/H] or [M/H].  Typically, surface gravity, log(g), is another physical parameter probed by spectroscopy.  In the case of M dwarf stars on the main sequence, log(g) is predicted to be a strict function of \teff and metallicity.  With \teff and metallicity alone, the stellar mass, radius and bolometric luminosity can be determined using either predictions from stellar evolutionary models \citep[e.g.][]{Dotter2008}, or empirical relations \citep[e.g.][]{Segransan2003,Boyajian2012}.

Moderate-resolution infrared spectra (R$\sim$2700, 1.5 to 2.5 $\mu$m) for {\it Kepler}-445 and {\it Kepler}-446 were obtained by \citet{Muirhead2014}, who determined stellar parameters based on the K-band indices of \citet{Rojas2012}, then interpolated those values on a new set of models based on the Dartmouth Stellar Evolution Database \citep{Dotter2008}, calculated by G. Feiden.  A visible low-resolution spectrum for {\it Kepler}-445 (R$\sim$900, 3200 to 9200\,$\angstrom$) was obtain by \citet{Mann2013c}, who used a new calibration to measure effective temperature.  They then used the mass-radius-temperature relations of \citet{Boyajian2012} for their sample, but were unable to determine the stellar mass and radius of {\it Kepler}-445 as it was too cool for these relations.  {\it Kepler}-446 was not discovered by the {\it Kepler} pipeline at the time of their study.

In this paper, we combine the techniques from  \citet{Mann2013c} and \citet{Muirhead2014} to produce the best possible physical parameters of the stars.  We choose to use the \citet{Mann2013c} visible-light spectroscopic effective temperature calibration, since it is calibrated using truly empirical measurements of nearby M dwarf stars from optical-long baseline interferometry \citep{Boyajian2012}.  We use the K-band spectroscopic metallicity measurements from \citet{Muirhead2014}, as those used empirical calibrations by \citet{Rojas2012}, who verified the metallicities by comparison to space-motions of nearby M dwarf stars.  Finally, we interpolate those values onto the new suite of Dartmouth evolutionary models as was done in \citet{Muirhead2014}, but with a small correction based on the empirical measurements of Barnard's Star \citep[similar to ][]{Muirhead2012a}.

To do this, we used primarily the archival spectroscopy described above, but acquired a new visible spectrum of {\it Kepler}-446.  Our additional spectrum of {\it Kepler}-446 was acquired with the SNIFS instrument on the University of Hawaii 88-inch Telescope at Mauna Kea Observatory.  We followed identical observing and reduction procedures described in \citet{Mann2013c}.

To stitch the visible and near-infrared spectra together for each star, we combine the $r$ magnitudes measured in the previous section with the $H$ and $Ks$ magnitudes from 2MASS.  We acquired the filter transmission curves for each band ($r$, $H$ and $K$) and computed a synethic spectral magnitude using the transmission curve and the measured spectrum, keeping in mind that 2MASS used a modified Vega system, and our measured SDSS magnitudes use the AB system \citep{Oke1983}.  We calculated the difference between the synthetic spectral magnitude and the measured magnitude from each respective survey, and used that difference to renormalize and combine the visible and near-infrared spectra.

\begin{figure*}[]
\begin{center}
\includegraphics[width=6.5in]{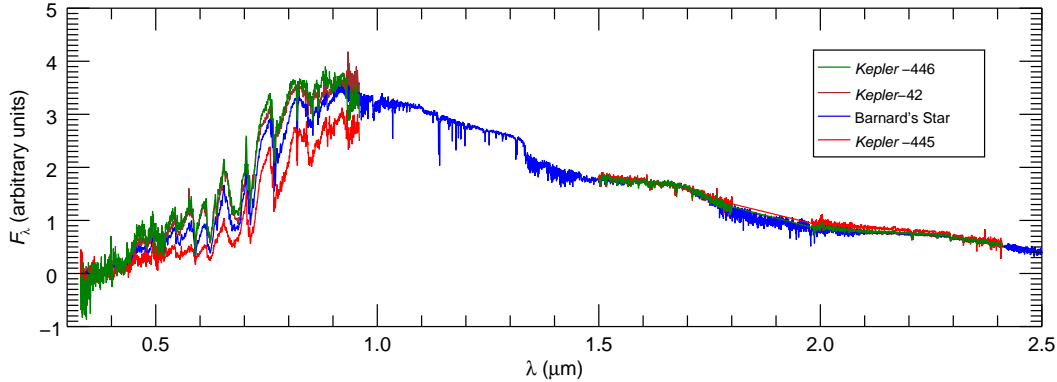}
\caption{Visible and near-infrared spectra of {\it Kepler}-445 and {\it Kepler}-446, with spectra of Barnard's Star and {\it Kepler}-42 for comparison, all normalized to match in H-band ($\lambda \sim 1.65$ $\mu$m).  The visible spectrum of {\it Kepler}-446 is part of this work, with the remaining visible and infrared spectra coming from \citet{Mann2013c} and \citet{Muirhead2014}.  The spectra were first renormalized piece-wise to match infrared photometry from the Two-Micron All Sky Survey  \citep[2MASS,][]{Cutri2003,Skrutskie2006} and our independent SDSS $r$ band photometry.  The similar shapes to the spectra strongly support our interpretation that {\it Kepler}-445 and {\it Kepler}-446 are mid-M dwarf stars with similar masses and radii to Barnard's Star and {\it Kepler}-42.  We have made no correction for interstellar reddening, as these stars are all expected to be within 150 pc of the sun (see Table \ref{stellar_table}).\label{spectra_figure}}
\end{center}
\end{figure*}

Figure~\ref{spectra_figure} plots the visible and infrared spectra of {\it Kepler}-445 and {\it Kepler}-446, with spectra of Barnard's Star from \citet{Mann2013c} and {\it Kepler}-42 from \citet{Muirhead2014} for comparison, subject to an arbitrarily total normalization such that they overlap in $H$-band.  Clearly, the spectra are very similar: are all identifiably M4 dwarf stars.  We did not correct for interstellar reddening because all stars are likely well within 150 pc of the sun (see Section~\ref{stellar_table}).  None of the stars show H$\alpha$ in emission, indicating low quiescent-activity and that all of the stars are likely old \citep[$t>5$ Gyr,][]{West2008}.  However, we note that Barnard's Star does show occasional H$\alpha$ flaring \citep[e.g.][]{Paulson2006}, and that age-activity relationships are a statistical tool for estimating the age of an ensemble of stars, and are subject to large uncertainties when applied to any specific star.

\begin{deluxetable*}{lcccccc}
\tablewidth{0pt}
\tablecaption{Stellar Parameters$^{\dagger\dagger}$\label{stellar_table}}
\tablehead{
\colhead{Mid-M Dwarf Star} &
\colhead{\teff} & 
\colhead{[Fe/H]} &
\colhead{[M/H]} &
\colhead{$M_\star$} &
\colhead{$R_\star$} &
\colhead{Distance}
\
}
\startdata
{\it Kepler}-445 &
3157 $\pm$ 60 K\tablenotemark{a} & 
+0.27 $\pm$ 0.13\tablenotemark{c} & 
+0.19 $\pm$ 0.12\tablenotemark{c} & 
0.18 $\pm$ 0.04 $M_\sun$ & 
0.21 $\pm$ 0.03 $R_\sun$ & 
$\sim$ 90 pc \\

{\it Kepler}-446 &
3359 $\pm$ 60 K & 
-0.30 $\pm$ 0.12\tablenotemark{c} & 
-0.21 $\pm$ 0.12\tablenotemark{c} & 
0.22 $\pm$ 0.05 $M_\sun$ & 
0.24 $\pm$ 0.04 $R_\sun$ & 
$\sim$ 120 pc \\

{\it Kepler}-42 &
3241 $\pm$ 57 K\tablenotemark{a} & 
-0.48 $\pm$ 0.12\tablenotemark{d} & 
-0.33 $\pm$ 0.12\tablenotemark{d} & 
0.15 $\pm$ 0.03 $M_\sun$\tablenotemark{d} & 
0.18 $\pm$ 0.02 $R_\sun$\tablenotemark{d} & 
$\sim$ 40 pc\tablenotemark{d} \\

Barnard's Star &
3238 $\pm$ 11 K\tablenotemark{a} & 
-0.39 $\pm$ 0.17\tablenotemark{c} &
-0.27 $\pm$ 0.12\tablenotemark{c} &
0.159 $\pm$ 0.013 $M_\sun$\tablenotemark{a} &
0.1867 $\pm$ 0.0012 $R_\sun$\tablenotemark{e} &
1.824 $\pm$ 0.005 pc\tablenotemark{f}

\enddata
\tablenotetext{$^{\dagger\dagger}$}{Values without a mark are from this work, values with a mark are from the following: (a) \citet{Mann2013c}, (b) \citet{Rojas2012}, (c) \citet{Muirhead2014}, (d) \citet{Muirhead2012a}, (e) \citet{Boyajian2012}, (f) {\it Hipparcos} astrometric parallax from \citet{vanleeuwen2007}.}
\end{deluxetable*}

To determine any offsets to apply to the new Dartmouth models, we interpolated the empirical effective temperature and metallicity measurements of Barnard's Star onto the new Dartmouth models to determine the predicted stellar mass and radius.  The predicted mass and radius are marginally  different from the empirical mass \citep[determined using mass-luminosity relationships of][]{Delfosse2000} and the empirical radius \citep[measured using optical long-baseline interferometry,][]{Lane2001,Boyajian2012}, with offsets of  -0.004 $Msun$ and 0.002 $Rsun$, both of which are well below the typical uncertainty in those quantities.  Nevertheless we still applied those corrections to the interpolated values.

Our results for the stellar parameters appear in Table \ref{stellar_table}.  In addition to calculating the parameters for {\it Kepler}-445 and {\it Kepler}-446, we also recalculated the parameters for {\it Kepler}-42 using identical methods, and find agreement with \citet{Muirhead2012a}, who used a similar technique, but applied a correction to the original Dartmouth models \citep{Dotter2008}, rather than the new set used in \citet{Muirhead2014} and this study.  We estimated the distance to the stars by inverting the mass-luminosity relations of \citet{Delfosse2000} to determine the stars absolute $K$-band magnitudes, and compared that to the measured $K$-band magnitude from 2MASS.  All three stars are relatively nearby with distances less than 150 parsecs.

The stellar parameter determinations for {\it Kepler}-445 and {\it Kepler}-446 presented here revise values published in the literature.  Compared to \citet{Muirhead2014}, the masses and radii presented here are slightly larger owing to the use of the \citet{Mann2013c} effective temperature calibration rather than the K-band technique.  The most recent evaluation of stellar parameters for {\it Kepler} targets was compiled by \citet{Huber2014}, who assign both {\it Kepler}-445 and {\it Kepler}-446 smaller and larger radii (respectively) than presented here.  We ascribe this to erroneous metallicity determinations in that catalog, which lists {\it Kepler}-445 as a metal-poor star ([Fe/H] = -0.380) and {\it Kepler}-446 as a metal-rich star ([Fe/H]=+0.30), the inverse of our measurements.  The source for the {\it Kepler}-446 parameters in \citet{Huber2014} is \citet{Dressing2013}, who used photometry to estimate stellar parameters.  As \citet{Dressing2013} admit, photometry alone places poor constraints on M dwarf metallicity, and {\it Kepler}-446's erroneously large radius in both \cite{Dressing2013} and \cite{Huber2013} is a direct consequence of the imprecise metallicity measurement.

\subsection{High-resolution Spectra and Space Motions}

We also acquired higher-resolution spectra in order to measure the absolute radial velocities of the stars and their galactic space motions.  We observed {\it Kepler}-445 and {\it Kepler}-446 with the Echellette Spectrograph and Imager (ESI) on the Keck-II telescope \citep{Sheinis2002} on UT 25 July 2014.  We operated ESI in echellette mode using a slit width of 0.5 arcseconds, achieving a resolving power of $\sim$ 8000 from 4000 to 10000 $\angstrom$, spread across ten cross-dispersed orders.  For {\it Kepler}-446 we acquired a single 1000 second exposure, and for {\it Kepler}-445 we co-added three 900 second exposures.  We also observed GJ 687 and GJ 905 as M-dwarf radial velocity standard stars with published radial velocities and uncertainties in \citet{Nidever2002} and \citet{Deshpande2012}, respectively.  We reduced the data using the publicly available ESIRedux pipeline\footnote{http://www2.keck.hawaii.edu/inst/esi/ESIRedux/index.html}, using dome flats for flat-fielding and arc lamps for wavelength calibration \citep{Prochaska2003,Bochanski2009}.

We then cross-correlated the radial velocity standard spectra with {\it Kepler}-445 and {\it Kepler}-446 to measure their absolute radial velocities.  We used the full-width-at-half-maximum of the peak in the cross-correlation function as our measurement uncertainty, combined in quadrature with the archival measurement errors for the radial velocity standard stars.  We found agreement within our uncertainties when using GJ 687 or GJ 905 separately as standards, and we report the mean of those measurements here.  For {\it Kepler}-445 we measured an absolute radial velocity of -61 $\pm$ 1 km s$^{-1}$, and for {\it Kepler}-446 we measured -118 $\pm$ 1 km s$^{-1}$.

When combined with archival proper motion measurements and our distance estimates from the previous section, we can estimate the stars' space motions.  The most recent proper motion measurements available for {\it Kepler}-445 and {\it Kepler}-446 are from the PPMXL catalog \citep{Roeser2010}.  They report a proper motion of $\mu_\alpha$=42.2 $\mu_\delta$=132.7 mas yr$^{-1}$ for {\it Kepler}-445 and $\mu_\alpha$=-13.2 $\mu_\delta$=-30.6 mas yr$^{-1}$ for {\it Kepler}-446.  Both proper motion measurements are relatively large and are consistent with M dwarf stars within 150 pc.

\begin{figure}[]
\begin{center}
\includegraphics[width=3.5in]{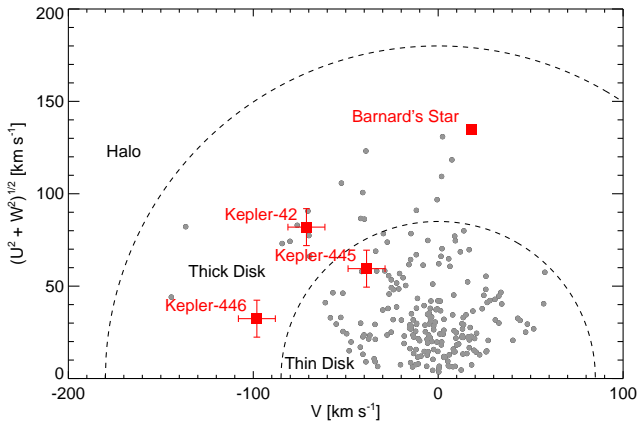}
\caption{Toomre Diagram of G K and M type stars with measured trigonometric parallaxes greater than 100 $mas$, the same stars used in a similar figure in \citet[][, Fig. 9]{Muirhead2012a}.  We include the thin disk, thick disk and halo boundaries from \citet{Fuhrmann2004} for reference.  Barnard's Star, {\it Kepler}-42, {\it Kepler}-445 and {\it Kepler}-446 are all shown.  {\it Kepler}-42 and {\it Kepler}-446 are located in the thick disk regime, consistent with their low metallicities, whereas {\it Kepler}-445 is located in the thin disk regime, consistent with its high metallicity.  We corrected for the solar motion using the values of \citet{Coskunoglu2011}:U = -8.5, V = 13.38, and W = 6.49 km/s.  We note that the distances to {\it Kepler}-42, {\it Kepler}-445 and {\it Kepler}-446 and their corresponding tangential motions are highly uncertain, as is the true solar motion through the galaxy.\label{toomre_figure}}
\end{center}
\end{figure}

Combining proper motion, radial velocity and distance estimates, we measure the galactic space velocity motions of {\it Kepler}-445 to be U = 59, V = -39, W = 9 km s$^{-1}$, and for {\it Kepler}-446 we measure U = 8, V = -98, W = -31 km s$^{-1}$.   We corrected for the galactic solar motion using the values of \citet{Coskunoglu2011}:U = -8.5, V = 13.38, and W = 6.49 km/s.  Figure~\ref{toomre_figure} plots the space motions for {\it Kepler}-445, {\it Kepler}-446, {\it Kepler}-42 and Barnard's Star with nearby dwarf stars in a Toomre diagram with thin disk, thick disk and halo boundaries from \citet{Fuhrmann2004}.  {\it Kepler}-42 and {\it Kepler}-446 are located in the thick disk regime, consistent with their low metallicities, whereas {\it Kepler}-445 is located in the thin disk regime, consistent with its high metallicity.  We note that the distances to {\it Kepler}-42, {\it Kepler}-445 and {\it Kepler}-446 and their corresponding tangential motions are highly uncertain, as is the true solar motion through the galaxy.  Therefore, we estimate the uncertainty in their galactic space motions to be $\sim$10 km s$^{-1}$.

\subsection{Kepler Photometry}

{\it Kepler}-445 was observed by {\it Kepler} in quarters 6, 8 and 9 in long-cadence mode as part of guest observer program GO20031, a search for microlensing by observing high-proper motion stars \citep{DiStefano2010}.  It was later observed in quarters 12, 13, 14 and 16 in long-cadence mode and in quarter 17 in short-cadence mode as an exoplanet search target as part of the primary {\it Kepler} Mission.  However, quarter 17 was cut short due to a malfunction in the spacecraft ending the primary mission, so we did not include it in our analysis.

{\it Kepler}-446 was observed by {\it Kepler} in quarter 7 in long-cadence mode as part of guest observer program GO20001, a search for eclipsing binary systems in a sample of M dwarf stars \citep{Harrison2010}.  It was again observed in quarters 12, 13, 14 15 and 16 in long-cadence mode and in quarter 17 in short-cadence mode as an exoplanet search target as part of the primary {\it Kepler} Mission, though we do not include the quarter 17 data for the reason stated above.

\subsubsection{Known Transiting Planets}

At the time this manuscript was submitted, the NASA Exoplanet Archive \citep[NEA,][]{Akeson2013} listed two planets orbiting {\it Kepler}-445 (KOI-2704), with orbital periods ($P$) of 4.871229 $\pm$ 1.1e-05 and 2.984151 $\pm$ 1.1e-05 days, for {\it Kepler}-445c and {\it Kepler}-445b respectively (KOI-2704.01 and KOI-2704.02), with planetary radii ($R_P$) of 2.9 and 1.96 $R_\earth$, respectively.  The NEA listed three planets orbiting {\it Kepler}-446 (KOI-2842) with orbital periods of 1.5654088 $\pm$ 3.3e-06, 5.148921 $\pm$ 2.2e-05, 3.03617925 $\pm$ 5.49e-06 days for {\it Kepler}-446b, {\it Kepler}-446d and {\it Kepler}-446c respectively (KOI-2842.01, KOI-2842.02 and KOI-2842.03), with planetary radii ($R_P$) of 25 $\pm$ 15, 26 $\pm$ 15 and 2.393 $\pm$ 0.582 $R_\earth$.  For {\it Kepler}-446, the NEA listed notably large impact parameters ($b$) of of 1.17 $\pm$ 0.92, 1.19 $\pm$ 0.99 and 0.90  $\pm$ 0.177, for {\it Kepler}-446b, {\it Kepler}-446d and {\it Kepler}-446c respectively.  

The planetary radii listed in the NEA as initially downloaded were determined using either an MCMC fitting routine, as in the case of {\it Kepler}-445c, {\it Kepler}-445b, {\it Kepler}-446b and {\it Kepler}-446d \citep{Burke2014}, or by the {\it Kepler} pipeline when initially searching for threshold-crossing events, as in the case of {\it Kepler}-446c \citep{Tenenbaum2014}, with the stellar parameters from \citet{Huber2014}.  For {\it Kepler}-446b and {\it Kepler}-446d We attribute the large radii and uncertainties in the NEA to correlations between the transit impact parameter and planet-to-star radius ratio.  The parameters are typically not strongly correlated; however in the case of {\it Kepler}-446 the transit durations ($T$) are relatively short, lasting only about an hour each, and the long-cadence integrations times are $\sim$30 minutes.  This results in the light-curve being significantly smoothed and appearing to be V-shaped, where a truly flat-bottomed, low-impact parameter transit curve is indistinguishable from a grazing eclipse.

We note that when this manuscript was accepted, the parameters listed in the NEA had changed, listing more reasonable values for the planet radii for {\it Kepler}-446, but still significantly larger than the values we calculate in Section \ref{transit_fits}.

\subsubsection{Independent Planet Search}

The {\it Kepler}-445 and {\it Kepler}-446 transiting planet-candidates found by the {\it Kepler} pipeline did not utilize all quarters of data available currently.  Therefore, we conducted our own search for additional transiting planets on the complete, available \Kepler\ dataset for {\it Kepler}-445, {\it Kepler}-446 and {\it Kepler}-42.  We downloaded the {\it Kepler} light curves for {\it Kepler}-445 and {\it Kepler}-446 from the NASA's Mikulski Archive for Space Telescopes (MAST).  The light curve data files contain flux values measured using simple aperture photometry on the {\it Kepler} pixels using pre-defined apertures (\texttt{SAP\_FLUX}), and flux values that have been detrended using custom tools to remove slowly-varying instrumental fluctuations in the {\it Kepler} photometric response as well as fluctuations due to stellar rotation and intrinsic variability: the ``pre-search data conditioning simple aperture photometry flux''  \citep[\texttt{PDCSAP\_FLUX}.][]{Smith2012}.  We choose to use the \texttt{PDCSAP\_FLUX} for the independent planet search and for fitting the transit events.

For each star, we fitted the \texttt{PDCSAP\_FLUX} light curve with a cubic basis spline (cubic B-spline) with breakpoints located 1.5 days apart, and divided the light curve by the B-spline fit to remove stellar and instrumental variability. We excluded outlier data points (both astrophysical and otherwise) from the B-spline fit by iteratively fitting the B-spline, locating points falling more than 3$\sigma$ away from the fit, and re-calculating the B-spline while excluding the outliers. We repeated this process until convergence, typically 5 iterations. We then searched for transits by calculating a Box Least Squared periodogram \citep[BLS,][]{Kovacs2002} for each star. We evaluated the BLS periodogram over periods ranging from 0.15 days (or 3.6 hours) to the total duration of the \Kepler\ observations. We evaluated the BLS power spectrum at roughly $10^5$ to $10^6$ discrete periods, depending on the total time baseline of Kepler observations, and we spaced the trial periods to ensure that the BLS signal of a short duration transit around a mid M-dwarf would not be smeared out by coarse period spacing. After we calculated the BLS power spectrum, we subtracted away a noise floor from the power spectrum and estimated its typical scatter by calculating the Median Absolute Deviation (MAD) and dividing by 0.67 to convert to an equivalent standard deviation. We considered any peak with a BLS signal-to-noise ratio of greater than 9 to be significant. Upon detecting a significant signal, we masked out the signal in question and re-calculated the periodogram.

\begin{figure}[]
\begin{center}
\includegraphics[width=3.3in]{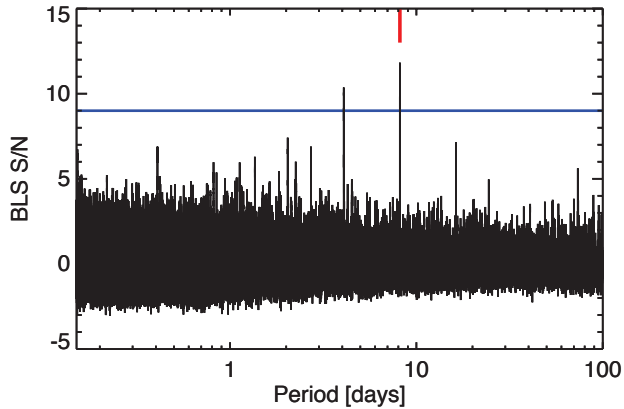}
\caption{BLS periodogram of {\it Kepler}-445 with the two planets detected by the \Kepler\ pipeline removed. There are a series of peaks with power greater than our threshold (horizontal blue line) corresponding to the true period (denoted by the vertical red hash mark) and harmonics of the new planet {\it Kepler}-445d.\label{bls}}
\end{center}
\end{figure}

In all three systems, we recovered the planets discovered by the \Kepler\ pipeline at high significance. For {\it Kepler}-42 and {\it Kepler}-446, after removing the three signals detected by the Kepler pipeline, there remained no significant peaks in the BLS, but for {\it Kepler}-445, after removing the two known planet candidates, there remained a series of significant peaks spaced like harmonics of a transiting planet signal, shown in Figure \ref{bls}. The highest peak in the BLS spectrum indicated a candidate period of 8.15275 days. Hereafter, we refer to this new planet candidate as {\it Kepler}-445d, orbiting near a 5:3 resonance with {\it Kepler}-445c, and we confirm all of the planet in Section \ref{sec:fpp}.  We also detected no significant photocenter shift in the centroid of the {\it Kepler} image of {\it Kepler}-445 during the {\it Kepler}-445d transit events.  Taking the difference between the centroid of the image in and out of transit, we calculate a photocenter shift of 0.00054 $\pm$ 0.00033 arcseconds.

\subsubsection{Transit Fits}\label{transit_fits}

\begin{deluxetable*}{lcccccc}
\tablewidth{0pt}
\tablecaption{Transit Parameters for {\it Kepler}-445}

\tablehead{
    \colhead{Parameter}
   & \colhead{{\it Kepler}-445b}
   & \colhead{{\it Kepler}-445c}
   & \colhead{{\it Kepler}-445d}
}
\startdata

$P$ (days)\tablenotemark{a} & 
2.984151 $\pm$ 0.000011 &
4.871229 $\pm$ 0.000011 &
8.15275 $\pm$ 0.00040 \\

$t_0$ - 2454833 (BJD) \tablenotemark{a} & 
133.1194 $\pm$ 0.0033 &
133.6408 $\pm$ 0.0019 &
3.7512226 $\pm$ 0.05 \\

$R_P/R_\star$  & 
0.0676 $\pm$ 0.0018 &
0.1075 $\pm$ 0.0014 &
0.0533 $\pm$ 0.0029 \\

$b$ & 
0.001 $^{+0.203}_{-0.001}$ &
0.000 $^{+0.101}_{-0.000}$ &
0.011 $^{+0.470}_{-0.010}$ \\

$T$ (hours) & 
1.0304 $\pm$ 0.0180 &
1.2287 $\pm$ 0.0154 &
1.3860 $\pm$ 0.0888 \\

$\mu_1$ \tablenotemark{b} & 
0.141 $\pm$ 0.080 &
0.141 $\pm$ 0.080 &
0.141 $\pm$ 0.080 \\

$\mu_2$ \tablenotemark{b} & 
0.263 $\pm$ 0.071 &
0.263 $\pm$ 0.071 &
0.263 $\pm$ 0.071 \\

$e$ \tablenotemark{a} & 
0 &
0 &
0 \\

\\ \hline \hline \\

Inc (degrees)\tablenotemark{c} & 
89.74 $^{+0.18}_{-0.28}$ &
89.91 $^{+0.07}_{-0.10}$ &
89.61 $^{+0.27}_{-0.25}$ \\

$a/R_\star$\tablenotemark{c} & 
21.94 $\pm$ 0.30 &
30.21 $\pm$ 0.38 &
42.95 $\pm$ 0.58 \\ 

\\ \hline \hline \\

$R_P$ ($R_\Earth$)\tablenotemark{c} & 
1.58 $\pm$ 0.23 &
2.51 $\pm$ 0.36 &
1.25 $\pm$ 0.19 \\

$S / S_0$\tablenotemark{d} & 
8.57 $\pm$ 0.60 &
4.52 $\pm$ 0.31 &
2.24 $\pm$ 0.17 \\

$M_P$ ($M_\earth$)\tablenotemark{e} & 
4 to 6 &
8 to 9 &
3 to 4 \\

$K$ (m s$^{-1}$)\tablenotemark{e} & 
6 to 8 &
9 to 10 &
2 to 4 \\

\enddata
\tablenotetext{a}{Held fixed in fitting procedure.  Periods and ephemerides for {\it Kepler}-445c and {\it Kepler}-445b are from the NASA Exoplanet Archive \citep{Akeson2013}.}
\tablenotetext{b}{Limb-darkening coefficients were tied between planets, and subject a prior based on \citet{Claret2011}.}
\tablenotetext{c}{Calculated from parametrization.}
\tablenotetext{d}{$S_0$ = 1360 W m$^{-2}$, the solar flux incident on Earth's upper atmosphere.}
\tablenotetext{e}{Coarsely estimated using the empirically-measured planet mass-radius relationships of \citet{Marcy2014}.}
\label{tbl:tapmcmc1}
\end{deluxetable*}

\begin{deluxetable*}{lcccccc}
\tablewidth{0pt}
\tablecaption{Transit Parameters for {\it Kepler}-446}

\tablehead{
    \colhead{Parameter}
   & \colhead{{\it Kepler}-446b}
   & \colhead{{\it Kepler}-446c}
   & \colhead{{\it Kepler}-446d}
}
\startdata

$P$ (days)\tablenotemark{a} & 
1.565409 $\pm$ 0.0000033&
3.036179 $\pm$ 0.0000055&
5.148921 $\pm$ 0.000022  \\

$t_0$ - 2454833 (BJD) \tablenotemark{a} & 
132.9135 $\pm$ 0.0019 &
134.069573 $\pm$ 0.000945 &
133.3196 $\pm$ 0.0042 \\

$R_P/R_\star$  & 
0.0574 $\pm$ 0.0026 &
0.0424 $\pm$ 0.0018 &
0.0519 $\pm$ 0.0022 \\

$b$ & 
0.601 $^{+0.096}_{-0.088}$ &
0.025 $^{+0.529}_{-0.025}$ &
0.705 $^{+0.057}_{-0.066}$ \\

$T$ (hours) & 
0.6456 $\pm$ 0.0456 &
0.9432 $\pm$ 0.0504 &
0.8832 $\pm$ 0.0480 \\

$\mu_1$ \tablenotemark{b} & 
0.447 $\pm$ 0.059 &
0.446 $\pm$ 0.059 &
0.446 $\pm$ 0.059 \\

$\mu_2$ \tablenotemark{b} & 
0.353 $\pm$ 0.065 &
0.353 $\pm$ 0.065 &
0.353 $\pm$ 0.065 \\

$e$ \tablenotemark{a} & 
0 &
0 &
0 \\

\\ \hline \hline \\

Inc (degrees)\tablenotemark{c} & 
87.42 $^{+0.62}_{-0.37}$ &
88.97 $^{+0.57}_{-0.46}$ &
88.72 $^{+0.17}_{-0.19}$ \\

$a/R_\star$\tablenotemark{c} & 
14.20 $\pm$ 0.94 &
22.40 $\pm$ 1.36 &
31.60 $\pm$ 2.08 \\

\\ \hline \hline \\

$R_P$ ($R_\Earth$)\tablenotemark{c} & 
1.50 $\pm$ 0.25 &
1.11 $\pm$ 0.18 &
1.35 $\pm$ 0.22 \\

$S / S_0$\tablenotemark{d} & 
26.22 $\pm$ 4.70 &
10.54 $\pm$ 2.03 &
5.30 $\pm$ 0.83 \\

$M_P$ ($M_\earth$)\tablenotemark{e} & 
4 to 5 &
2 to 4 &
3 to 5 \\

$K$ (m s$^{-1}$)\tablenotemark{e} & 
6 to 8 &
2 to 5 &
3 to 5 \\

\enddata
\tablenotetext{a}{Held fixed in fitting procedure.  Periods and ephemerides are from the NASA Exoplanet Archive \citep{Akeson2013}.}
\tablenotetext{b}{Limb-darkening coefficients were tied between planets, and subject a prior based on \citet{Claret2011}.}
\tablenotetext{c}{Calculated from parametrization.}
\tablenotetext{d}{$S_0$ = 1360 W m$^{-2}$, the solar flux incident on Earth's upper atmosphere.}
\tablenotetext{e}{Coarsely estimated using the empirically-measured planet mass-radius relationships of \citet{Marcy2014}.}
\label{tbl:tapmcmc2}
\end{deluxetable*}

Given the large uncertainties on the physical radii of the planets orbiting {\it Kepler}-446, and the new detection of {\it Kepler}-445d, we chose to fit the the transit light curves for {\it Kepler}-445 and {\it Kepler}-446 independently.  We fitted transit light curves to the data using a modified version of the Transit Analysis Package, or TAP, an IDL software package developed by \citet{Gazak2012}.  TAP employs a Markov-Chain Monte Carlo algorithm within a Bayesian framework for determining transit parameters.  In order to reduce the degeneracy between impact parameter and planet-to-star radius ratio, we modified TAP to impose a strict prior on the stellar densities for {\it Kepler}-445 and {\it Kepler}-446.  As \citet{Seager2003} showed, ingress/egress duration and full transit duration can be combined in such a way so as to determine the density of the host star, assuming knowledge of the planet's orbital period, orbital eccentricity and longitude of periastron.  Likewise, knowledge of the host star density, orbital eccentricity and longitude of periastron can be combined to constrain the relationship between ingress/egress duration and transit duration.  In the case of the planets orbiting {\it Kepler}-445 and {\it Kepler}-446, ingress/egress duration is difficult to measure due to the V-shaped nature of the light curves as discussed above, and imposing such a constraint is a powerful way to fit accurate transit parameters.  Indeed, \citet{Muirhead2012a} used this technique to fit the transit light curves for the {\it Kepler}-42 system, although they fixed the stellar density rather than applying a prior.

We assumed a stellar density prior based on the measured masses and radii for {\it Kepler}-445 and {\it Kepler}-446 from spectroscopy described in the previous section.  Due to the use of evolutionary models, the mass and radius uncertainties are nearly 100\% covariant for the stars, so we only use the uncertainty in mass when calculating the density prior.  We assumed a Gaussian prior with a mean of 26.51 gm cm$^{-3}$ and a standard deviation of 6.54 gm cm$^{-3}$ for {\it Kepler}-445, and a mean of 22.73 gm cm$^{-3}$ and a standard deviation of 6.06 gm cm$^{-3}$ for {\it Kepler}-446.

We also assumed that all of the planets orbiting {\it Kepler}-445 and {\it Kepler}-446 have low eccentricity ($e$), and fix the value to zero in our fit.  Following \citet[][, Eq. 1]{Wu2002}, and assuming reasonable planet densities with tidal $Q_P'$ values of 100 to 10000, all of the planets orbiting {\it Kepler}-445 and {\it Kepler}-446 have circularization timescales of less than 1 Gyr.  Given the lack of quiescent H$\alpha$ or $u$-band emission in either {\it Kepler}-445 or {\it Kepler}-446, and {\it Kepler}-446's high galactic space motion, we are confident that the stars are older than 1 Gyr and likely older than 5 Gyr.  We note, however, that planets can sustain eccentricities over long timescales via spin-orbit coupling, and that in multiple-planet systems non-zero eccentricity can be sustained via orbital resonances.  The assumption of zero eccentricity also negates the importance of the longitude of periastron parameter for the transit fit or for the imposed prior on stellar density.

For limb-darkening, we fit a linear ($u1$) and quadratic ($u2$) coefficient to the transits.  For each system, {\it Kepler}-445 and {\it Kepler}-446, we tied the limb-darkening parameters across each of the transiting planet fits.  We used Gaussian priors for the limb-darkening coefficients, based on the expected coefficients from \citet{Claret2011} using our measured stellar parameters and their uncertainties.  For {\it Kepler}-445 we used a Gaussian prior with a mean linear limb-darkening parameter of 0.50 with standard deviation of 0.17, and a quadratic limb-darkening parameter of 0.35 with a standard deviation of 0.13. For {\it Kepler}-446, we used a Gaussian prior with a mean linear limb-darkening parameter of 0.42 with standard deviation of 0.12, and a quadratic limb-darkening parameter of 0.35 with a standard deviation of 0.11.

\begin{figure}[]
\begin{center}
\includegraphics[width=3.5in]{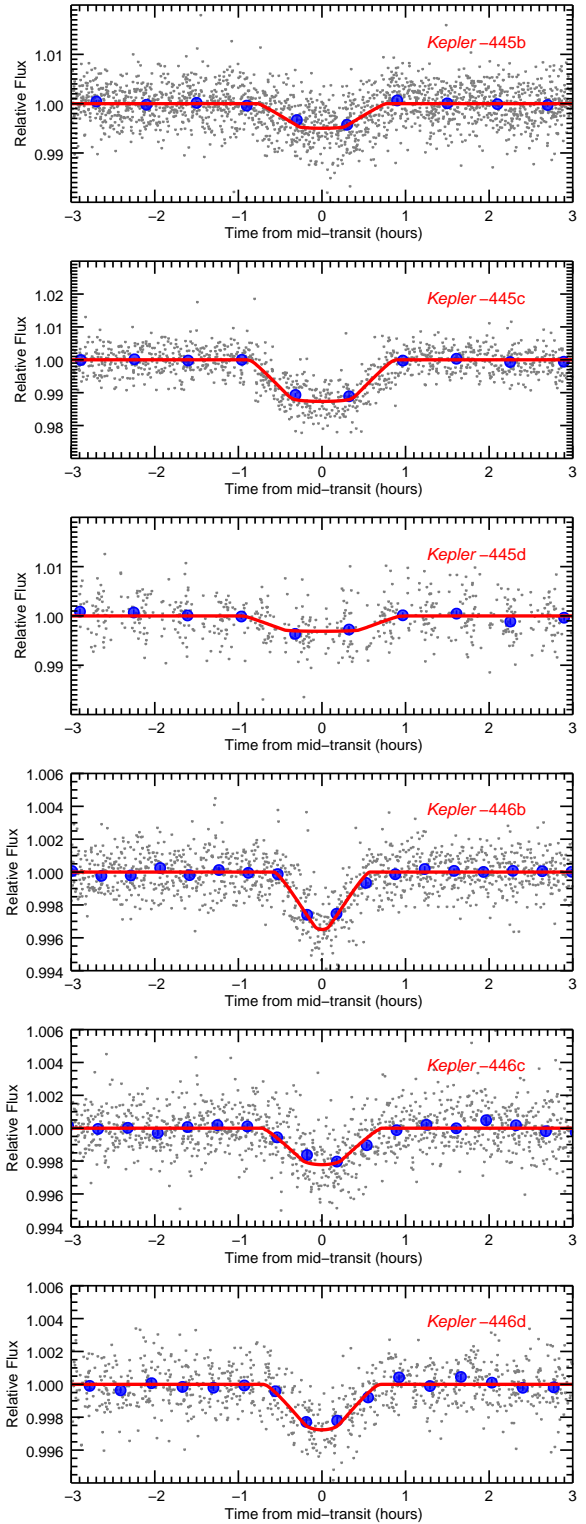}
\caption{Phase-folded light curves for the {\it Kepler}-445 and {\it Kepler}-446 transiting planets with our best fit transit light curves.  Raw data are shown as {\it grey dots}, binned data as {\it blue circles}, and the model as {\it red lines}.  Note the V-shaped transits for {\it Kepler}-446, which is due to the $\sim$30 minute integration times of the {\it Kepler} data.  We speculate that this contributed to the erroneous planet-to-star radius ratios returned by the {\it Kepler} pipeline.  Also note that the scale of the y-axis changes for the {\it Kepler}-445 and {\it Kepler}-446 plots.\label{transit_plot}}
\end{center}
\end{figure}

\begin{figure*}[]
\begin{center}
\includegraphics[width=6.8in]{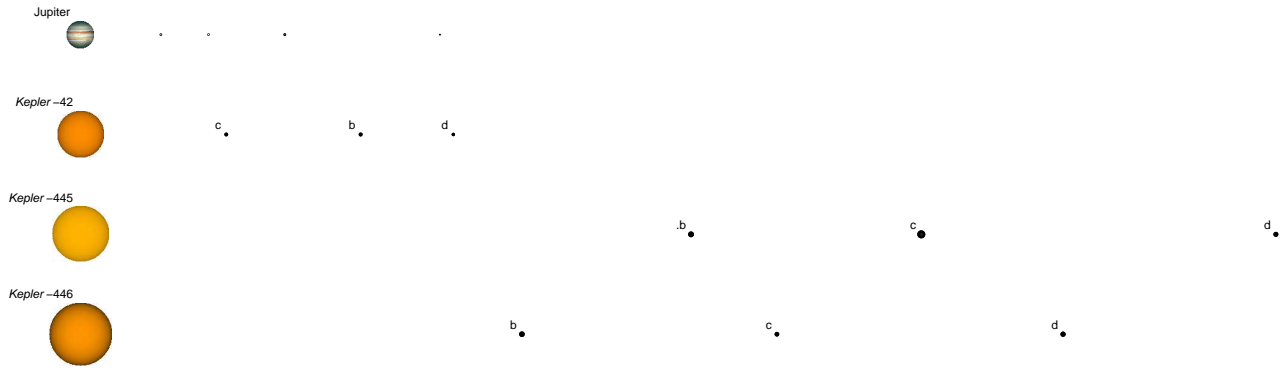}
\caption{Scaled illustration of the {\it Kepler}-42, {\it Kepler}-445 and {\it Kepler}-446 exoplanetary systems, with Jupiter and the Galilean moons shown for reference.  The colors of the stars match what is expected given the visible-light spectra shown in Figure \ref{spectra_figure}, and the limb-darkening applied to the images of the stars matches the measured limb-darkening from the transit curves.  The dramatic difference in the color and limb-darkening between {\it Kepler}-445 and {\it Kepler}-446/{\it Kepler}-42 is likely due to the markedly different metallicities, though we stress these are all M4 dwarf stars by spectral classification.  Interestingly, the higher metallicity of {\it Kepler}-445 results in a redder $r$-$J$ color, but a {\it bluer} appearance to the human eye compared to metal-poor stars, due to the peculiar wavelengths of deep molecular features within the visible spectrum (400 to 700 nm). \label{illustration}}
\end{center}
\end{figure*}

Finally, we chose to fix the orbital periods and transit epochs ($t_0$) of the planets to the values listed in the NEA for all but {\it Kepler}-445d, for which we fit a separate transit model to the \Kepler light curve using a Levenberg-Marquardt minimization routine.  We used the best fit period and transit epoch for the TAP fit.  Allowing the periods of the planets to vary within TAP had no significant effect on the resulting transit parameters.  Our final transit parameters for the six planets are listed in Tables \ref{tbl:tapmcmc1} and \ref{tbl:tapmcmc2}, and we show the phase-folded {\it Kepler} light curves and fitted transit curves in Figure~\ref{transit_plot}.  We combine the stellar parameters with the transit parameters to determine the physical parameters for the planets, which are also listed in Table \ref{tbl:tapmcmc1}, and Figure~\ref{illustration} illustrates the planets with {\it Kepler}-42 and the Galilean moons of Jupiter for comparison.  We also calculate the incident flux on the planets as a fraction of the solar flux incident on the Earth's upper atmosphere ($S_0$ = 1360 W m$^{-2}$).  The values indicate that all of the planets are likely too hot to be located within their host stars' habitable zones, using habitable-zone limits calculated by \citet{Kopparapu2013}.  However, with an $S / S_0$ of 2.24, one could argue that {\it Kepler}-445d is near the habitable zone.

We also include coarse estimates for the planet masses and expected semi-amplitude radial velocity signatures in Table \ref{tbl:tapmcmc1}, using the recent empirically-measured planet mass-radius relations of \citet{Marcy2014}.  All of the planets have anticipated radial velocity semi-amplitudes of over 1 m s$^{-1}$.  However, being mid-M dwarf stars, the stars are relatively faint for current-generation visible-light precision-radial-velocity spectrometers \citep[e.g.][]{Bottom2013}.  In the future, the stars may be compelling targets for next-generation, infrared precision-radial-velocity instruments \citep{Mahadevan2012, Halverson2014, Quirrenbach2012, spirou1, spirou2, spirou3, spirou4, spirou5, Crepp2014, Ge2014}.  Such measurements would prove useful for precisely measuring the planet masses and for placing constraints on their atmospheres \citep[e.g.][]{Kempton2012} and interior structures \citep[e.g.][]{Rogers2010,Valencia2013}, as well as measuring any non-zero eccentricity \citep[e.g.][]{Anglada2013}.

\section{False Positive Analysis}\label{sec:fpp}

As with the vast majority of \textit{Kepler} planet candidates, the transit signals detected in the light curves of {\it Kepler}-445 and {\it Kepler}-446 are not amenable to dynamical confirmation as \textit{bona fide} planets, either by radial velocity or transit timing variation measurements.  Confirming their planetary nature thus requires probabilistic validation; that is, demonstrating that the probability for them to be caused by a blended stellar eclipsing binary, or any other astrophysical false positive scenario, is very low.

While broad arguments have demonstrated that only a small number of \textit{Kepler} planet candidates are expected to turn out to be astrophysical false positives \citep{Morton2011,Fressin2013}, any specific planet candidates of particular interest should be individually investigated more thoroughly to ensure their planetary nature.  To this end, we have employed the method described in detail by \citet{Morton2012} to calculate false positive probabilities (FPPs) for all six of the candidates discussed in this paper.  This method compares the shapes of the observed light curves to the shapes of simulated planet and astrophysical false positive scenarios in order to determine the relative likelihoods of the signals to be caused by the different scenarios.  

The basic assumption of this analysis is that the observed transit signal is roughly spatially coincident with the target star---that is, that the flux decrement in the aperture of the target star is not due, for example, to a different star several arcseconds separated from the target star.  An important part of the \textit{Kepler} vetting pipeline is looking for such offsets \citep{Bryson2013}---only candidates that do not show significant offsets get promoted to ``planet candidate'' status.  The results of these tests are summarized in the NASA Exoplanet Archive tables in the column titled ``PRF $\Delta\theta_{\rm MQ}$'' and associated uncertainty, which indicates the difference in position between the \textit{Kepler}
pixel response function (PRF) fitted to the in- and out-of-transit data.  For {\it Kepler}-446b, c and d and {\it Kepler}-445b and d, all these offsets are significantly smaller than 1 arcsecond.  For {\it Kepler}-445d, which is not in the \textit{Kepler} catalogs, we perform our own in- and out-of-transit centroid analysis, finding no noticeable shift.  While we note that this analysis is not as detailed as the \citet{Bryson2013} PRF analysis, it certainly indicates that there is no large offset of the candidate signal.

While the pixel-level analysis of these candidates localizes them to be at least within 1 arcsecond of the target stars' locations, limiting the possible on-sky area in which could exist a potential false-positive blend, the constraints from the AO and aperture masking sensitivity curves presented in Section \ref{sec:ao} further restrict the parameter space of possible false positives.  Incorporating all these constraints into the \citet{Morton2012} analysis, we obtain FPPs for all candidates of $<10^{-4}$, except for {\it Kepler}-445d and {\it Kepler}-446d, for which we obtain FPPs of approximately 1 in 500 and 1 in 200, respectively.  For both {\it Kepler}-445d and {\it Kepler}-446d the most probable false positive scenario is an eclipsing binary within a hierarchical triple stellar system. 

We note, however, that these FPP calculations do not account for the fact that the candidates are observed to be in multiple-transiting systems.  As it is now known that many multiple planet systems have low mutual inclinations \citep{Fabrycky2012,Fang2012}, the presence of one transiting planet means that additional planets have a significantly higher probability to transit than if their orbital inclinations were randomly distributed.  Accounting for this effect gives a ``multiplicity boost'' to the probability of the planet scenario of about a factor of $\sim$5-10 (depending on the exact assumptions), thus lowering the FPPs of both {\it Kepler}-445d and {\it Kepler}-446d to below 0.1\%---all six of the planet candidates discussed in this paper are probabilistically validated.

However, there is another scenario, while not strictly an astrophysical false positive, that we cannot completely rule out.  That is, the AO and aperture masking observations presented here do not completely exclude the presence of stellar companions around {\it Kepler}-445 and {\it Kepler}-446.  This can be understood by noting that the inner working angle of 20 $mas$ achieved by the aperture masking observations corresponds to a projected distance of 2 AU at 100pc, and that a snapshot of a binary star system will generally not be at maximum separation, due to orbital position and inclination.  Using a Monte Carlo simulation of potential stellar companions distributed according to the stellar binary period and eccentricity distribution of \citet{Raghavan2010} (with randomized orbital elements), we find that about 25\% of potential stellar companions would be undetected by these high-resolution imaging observations.  However, a close stellar companion would be detrimental to the existence of the observed planets, so we place another constraint requiring putative stellar companions to have periods longer than 200 days; this removes another $\sim$15\% of potential stellar companions.  Thus, starting with an assumption that about 40\% of M-dwarf stars are in stellar binary systems \citep{Fischer1992, Clark2012, Duchene2013}, then our observations allow about a 4\% chance for either of {\it Kepler}-445 or {\it Kepler}-446 to be binary systems.

As discussed in \citet{Muirhead2012a}, there would be two potential consequences of the presence of such a stellar companion in either of these systems.  First, the planet radii would be larger than we estimate due to the unaccounted-for dilution (by a maximum of $\sqrt{2}$ if the planets are around the primary star, and by a larger factor if they are around a fainter secondary star).  Second, there is a possibility that the transiting planets could be distributed around both stars in the system, rather than all being around the same, single star.  Such a ``split-multi'' configuration, while predicted to be rare by \citet{Fabrycky2011}, does undoubtedly exist in the archetype KOI-284/Kepler-132 system \citep{Lissauer2014}, a visual binary with two of its three candidates having orbital periods of 6.18 and 6.42 days.  However, given the low probability for either of these two systems to be binaries, and the even lower probability that the planets would be split between the two components, we neglect this scenario for the remainder of this paper, although we do acknowledge its possibility at the $\sim$1\%-probability level.

\section{Occurrence of Compact Multiples Orbiting Mid-M Dwarf Stars}\label{occurrence}

With three confirmed compact multiple systems orbiting mid-M dwarf stars--{\it Kepler}-42, {\it Kepler}-445 and {\it Kepler}-446--we can estimate the occurrence of compact multiples like these around mid-M dwarf stars in general: defined here to be a nearly coplanar system of 2 or more planets all orbiting with periods of less than 10 days. First, we downloaded and combined the lists of all targets observed by {\it Kepler} for at least one quarter from quarter 1 to 16 from MAST \footnote{\url{http://archive.stsci.edu/pub/kepler/catalogs/}}.  The files include  measurements of the combined differential photometric precision \citep[CDPP,][]{Christiansen2012} for each target for each quarter, calculated over 3, 6 and 12 hour durations.

\begin{figure}[]
\begin{center}
\includegraphics[width=3.5in]{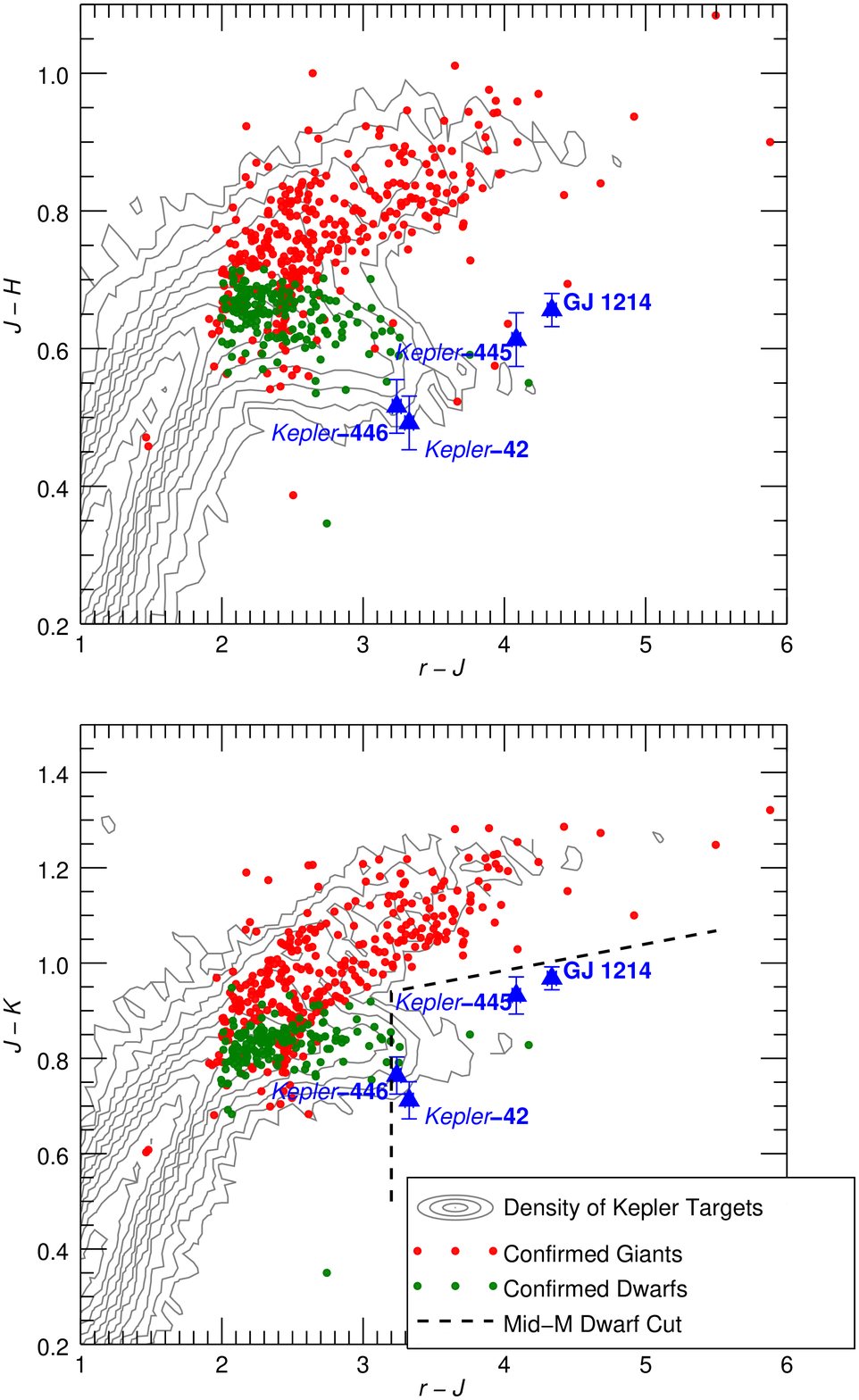}
\caption{{\it Top}: Density contours for KIC $J-H$ vs. $r-J$ of all objects observed by {\it Kepler} for at least one quarter from Q1 to Q16, with spectroscopically confirmed dwarf and giant stars from \citet[][, {\it green} and {\it red} circles, respectively]{Mann2012}.  {\it Bottom}: Same, but with $J$-$K$, which provides better dwarf-giant discrimination with no confirmed giant stars contaminating the dwarf sequence.  To calculate the occurrence of compact multiples orbiting mid-M dwarf stars, we isolate mid-M dwarf stars using a the color cuts indicated in the lower plot.  {\it Kepler}-42, {\it Kepler}-445 and {\it Kepler}-446 are shown ({\it blue triangles}), using our independent $r$ magnitude measurements.  We include GJ 1214, as it is the only other mid-M dwarf known to host a confirmed transiting planet \citet{Charbonneau2009}.  The lower positions of {\it Kepler}-446 and {\it Kepler}-42 compared to {\it Kepler}-445 and GJ 1214 in $J$-$H$ are consistent with their relative metallicities as measured by infrared spectroscopy \citep{Rojas2012, Muirhead2014}.  We note that many more mid-M dwarf stars were added in quarter 17, but this quarter was cut short and was not included in the current exoplanet search by the \Kepler pipeline.  Therefore, we did not include them in the occurrence calculation.\label{color_color_plot}}
\end{center}
\end{figure}

To calculate the occurrence of compact multiples orbiting mid-M dwarf stars, we use two approaches.  The first follows the methodology of \citet{Howard2012}, who calculated planet occurrence as a function of planet radius and orbital period using the {\it Kepler} planet candidate discoveries.  We considered the number of mid-M dwarf stars around which {\it Kepler} and the {\it Kepler} pipeline could have found compact multiples like the ones orbiting {\it Kepler}-42, {\it Kepler}-445 or {\it Kepler}-446.  We exclude {\it Kepler}-445d from our calculation, as that was not discovered by the {\it Kepler} pipeline.  We isolated mid-M-dwarf stars from the Kepler targets applying two color cuts using magnitudes from the KIC: $r-J>3.2$ to isolate red objects, and $J-K$ $<$ 0.0555 (r-J) + 0.7622 to remove evolved stars (giant and highly-reddened sub-giant stars).  At first we considered using $J-H$ to remove giant stars; however, we found that several giant stars identified in \citet{Mann2012} were not excluded using $J-H$, but were excluded using $J-K$, so we chose the latter.  We also excluded targets that were only observed in the very first quarter of {\it Kepler} (quarter 0), which was only 30 days and has lower intrinsic photometric precision than subsequent quarters, and targets that were added in quarter 17, since that has not yet been searched by the {\it Kepler} pipeline and was cut short due to a spacecraft malfunction.  The color cuts resulted in 509 mid-M dwarf stars with at least one full quarter of {\it Kepler} observations.  Figure~\ref{color_color_plot} illustrates the color-color cuts, and the position of {\it Kepler}-42, {\it Kepler}-445 and {\it Kepler}-446.  

Of the 509 mid-M dwarf stars, {\it Kepler}-42, {\it Kepler}-445 and {\it Kepler}-446 are the only stars with multi-planet-candidate systems listed on the NASA Exoplanet Archive at the time of this study.  In addition to these three stars, KOI-2862, KOI-4290, KOI-3855, KOI-3138 and KOI-5327 also host single transiting planet-candidates and fall within the defined color cuts.  We do not consider the single-transiting-planet hosts in this occurrence calculation.

We then calculated the number of mid-M dwarf stars with the signal-to-noise ratio necessary to discover each of the planets orbiting {\it Kepler}-42, {\it Kepler}-445 or {\it Kepler}-446.  To calculate the signal-to-noise ratio (S/N) for an arbitrary transiting planet orbiting a particular {\it Kepler} target, we used Equation A1 from \citet{Fressin2013}, repeated here with a slight modification:

\begin{equation}\label{snr}
{\rm S/N = {\delta \over CDPP_{eff}}}\sqrt{t_{\rm obs} \over P }
\end{equation}

where $\delta$ is the transit depth, $\rm CDPP_{eff}$ is the combined differential photometric precision over the duration of the transit, $t_{\rm obs}$ is the total {\it Kepler} observing time used to search for the planet by the {\it Kepler} pipeline, and $P$ is the planet orbital period.  To determine $\rm CDPP_{eff}$, we first averaged the 3, 6 and 12-hour CDPP values for each mid-M dwarf target over all quarters, then fit a function to the CDPP versus time with the form $\rm CDPP_{eff} = CDPP_0/\sqrt{t_{\rm dur}}$, where $t_{\rm dur}$ is the duration of a transit.  We then calculated the signal-to-noise that would be achieved on each of the planets orbiting {\it Kepler}-42, {\it Kepler}-445 (excluding {\it Kepler}-445d) and {\it Kepler}-446, if they were orbiting every single mid-M dwarf observed by {\it Kepler} with the same transit durations.

We adopt a value of 10 as the minimum signal-to-noise required to detect a planet, following \citet{Howard2012}.  This is a reasonable value, considering the signal-to-noise needed to detect the {\it Kepler}-42, {\it Kepler}-445 and {\it Kepler}-446 transit signals and give them a planet-candidates disposition.  For the three planets orbiting {\it Kepler}-42, they were detected by the {\it Kepler} pipeline using only the first quarter of data \citep{Borucki2011b}.  For {\it Kepler}-445 and {\it Kepler}-446, we combine the ``disposition provenance'' listed on the NASA Exoplanet Archive with the number of observed quarters within the provenance boundaries to determine the number of quarters required to detect their orbiting planets.  Following this, {\it Kepler}-445c and {\it Kepler}-445b were detected by the {\it Kepler} pipeline using two quarters of data (6 and 8).  The three planets orbiting {\it Kepler}-446 were detected by the {\it Kepler} pipeline using 1 quarter of data (7) for 0.01 and 0.02, and two quarters (7 and 12) for 0.03.  Table \ref{snr_table} indicates the resulting S/N values for the eight planets achieved when they were detected, all of which are greater than 10.  We also list the number of {\it Kepler} mid-M dwarf stars for which the S/N of the respective planet {\it would be} higher than 10, if it were transiting all of the Kepler mid-M dwarf stars.  Lastly, we list the individual transit probabilities ($R_\star / a$).

When calculating the number of mid-M dwarf targets for which {\it Kepler}-42, {\it Kepler}-445 or {\it Kepler}-446-like systems could be detected, we make two assumptions.  First, we assume the mid-M dwarf targets have similar radii to {\it Kepler}-42, {\it Kepler}-445 and {\it Kepler}-446, all roughly 0.20 $R_\sun$.  The true stellar radii of the mid-M dwarf targets affects the signal-to-noise of the {\it Kepler}-42, {\it Kepler}-445 or {\it Kepler}-446-like systems if there were transiting other stars.  The assumption of similar radii across all the mid-M dwarf targets is a reasonable one considering that {\it Kepler}-42, {\it Kepler}-445 and {\it Kepler}-446 themselves have similar radii and nearly subtend the full color-color space defined in the mid-M dwarf target cut (see Figure~\ref{color_color_plot}).  Even so, without spectra for the full sample of mid-M dwarf stars we must make this assumption.  Second, we assume that the impact parameter has a negligible impact on the signal-to-noise of the planet transits if they were transiting other mid-M dwarf targets.  In fact, an especially large impact parameter will reduce the signal-to-noise of a transit event since the duration will be shorter and the transit could even be grazing.

\begin{table}[]
\begin{center}
\caption{Planet S/N and Transit Probabilities\label{snr_table}
}
\begin{tabular}{c c c c}
\hline\hline
Planet & Transit S/N & {\it Kepler} Mid-Ms  & Transit \\
 & for disposition & where S/N  & probability \\
  & & would be $>$10 & \\
\hline
{\it Kepler}-42 b & 67.9 & 353 & 0.0826 \\
{\it Kepler}-42 c & 94.3 & 392 & 0.1587 \\
{\it Kepler}-42 d & 35.2 & 261 & 0.0621 \\
\hline
{\it Kepler}-445c & 26.3 & 468 & 0.0331 \\
{\it Kepler}-445b & 12.8 & 431 & 0.0456 \\
\hline
{\it Kepler}-446b & 24.5 & 412 & 0.0704 \\
{\it Kepler}-446d & 10.2 & 300 & 0.0316 \\
{\it Kepler}-446c & 14.5 & 301 & 0.0446 \\
    \hline 
 \end{tabular}
\end{center}
\end{table}

To calculate the occurrence of compact multiples, we use Equation 2 from \citet{Howard2012}, who calculated the frequency of planets for a specific range of orbital periods and planet-sizes, but modified for compact multiple systems:

\begin{equation}\label{fcell}
f_{\rm CM} = \sum_j^{n_{sys}} { 1 / p_j \over n_{sys,j} }
\end{equation}

Whereas \citet{Howard2012} calculated the number of planets per star for specific ranges of orbital period and planet radius, we, on the other hand, want to calculate the fraction of mid-M dwarfs that host compact multiple {\it systems}, which we call $f_{\rm CM}$.  Instead of indexing each {\it planet} detection in the summation, instead we index each {\it system}: $j$ refers to {\it Kepler}-42, {\it Kepler}-445 and {\it Kepler}-446, successively, $p_j$ refers to the transit probability of the whole compact multiple (corresponding to the transit probability of the outermost system), and $n_{sys,j}$ is the number of stars around which the system could be detected.  Equation \ref{fcell} implicitly defines a compact multiple as a system as being like {\it Kepler}-42, {\it Kepler}-445 and {\it Kepler}-446: 2 or more planets all orbiting with periods of less than 10 days.  

Also implicit in this formalism is the assumption that compact multiple systems are co-planar: if the outermost planet is detected, the inner planets would also be detected, presuming their signal-to-noise is high enough.  If, in fact, compact multiple systems are not coplanar, then $f_{\rm CM}$ represents a lower limit to their frequency around mid-M dwarf stars, as the single transiting-planet-candidate hosts metioned above could harbor many more interior planets that do not transit.

We calculated the uncertainty in $f_{\rm CM}$ following the same procedure as described in \citet{Howard2012}: We calculated the binomial probability distribution of drawing compact multiple planets from $n_{\rm pl} / f$ stars, and use the closest values to 1$\sigma$ above and below $f_{\rm CM}$ of the binomial distribution as the uncertainties ($\sim 15\%$ and $\sim 85\%$ of the cumulative binomial distribution).

Performing this calculation we arrive at a compact-multiple occurrence of 23 $^{+13}_{-7}$ \%.  The uncertainties correspond to drawing 5 and 2 compact multiple systems from 14 stars, based on the binomial distribution.  We note that we include mid-M dwarf stars of all metallicities observed by {\it Kepler} in this calculation, which roughly matches the metallicity distribution of the solar neighborhood \citep{Mann2013b}, and that {\it Kepler}-445 has significantly super-solar metallicity ([Fe/H]=+0.25) and {\it Kepler}-446 and {\it Kepler}-42 both have sub-solar metallicity ([Fe/H]$<$-0.30).  We conclude that roughly one-fifth to one-quarter of mid-M dwarf stars host compact multiple systems, for the full spectrum of metallicities in the solar neighborhood.

\subsection{A Monte Carlo Approach}

\begin{figure}[]
\begin{center}
\includegraphics[width=3.5in]{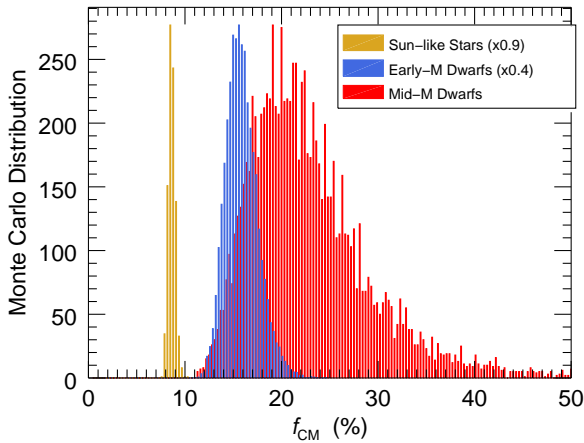}
\caption{Distribution of $f_{\rm CM}$ calculated for Sun-like stars, early M dwarf stars and mid-M dwarf stars, subject to arbitrary normalizations for clarity.  For mid-M dwarf stars, we simulated {\it Kepler}-42-, {\it Kepler}-445- and {\it Kepler}-446-like planets orbiting mid-M dwarf targets observed by {\it Kepler} for at least one full quarter, with random inclinations and subject to a randomized detectability.  $f_{\rm CM}$ was calculated using Equation \ref{fcm_equation}.  The distribution has a median of 26\%, a mode 21\% (with 1\% bins), and asymmetric uncertainties of $^{+7}_{-5}$ \%, corresponding to the equivalent of +/- 1$\sigma$, which is consistent with the value determined by Equation \ref{fcell}.  We performed identical calculations for Sun-like stars and early M dwarf stars, both of which show lower values of $f_{\rm CM}$ compared to mid-M dwarf stars, but the differences are less than 3$\sigma$.\label{fcm}}
\end{center}
\end{figure}

As a consistency check, we also performed the occurrence calculation using a Monte Carlo approach, taking into account the effect of transit duration on SNR, as well as the probability of a {\it Kepler}-pipeline detection for a given SNR.  We assigned each of the {\it Kepler} mid-M dwarf targets with at least one-quarter of data a random inclination, then calculated the SNR of the transit signals if the {\it Kepler}-42, {\it Kepler}-445 or {\it Kepler}-446 planets were transiting at the random inclination.  We then used the detectability ramp described in \citet{Fressin2013} to randomly determine whether each system would be detected by the {\it Kepler} pipeline.  This process results in the number of {\it Kepler}-42-, {\it Kepler}-445- and {\it Kepler}-446-like systems that should have been detected if they were orbiting all mid-M dwarf stars.  Since one {\it Kepler}-42, one {\it Kepler}-445 and one {\it Kepler}-446 system was detected, we use the following summation to determine $f_{\rm CM}$:

\begin{equation}\label{fcm_equation}
f_{\rm CM} = { 1 \over n_{Kepler-42} } + { 1 \over n_{\rm {\it Kepler}-445} } + { 1 \over n_{\rm {\it Kepler}-446} }
\end{equation}

\noindent where $n$ refers to the number of systems that would have been detected if they orbited all mid-M dwarf stars.  We repeated the simulation 10000 times, and the resulting distribution of $f_{\rm CM}$ is shown in Figure \ref{fcm}.  The distribution has a median of 22\%, a mode 21\%, and asymmetric uncertainties of $^{+7}_{-5}$ \%, corresponding to the equivalent of +/- 1$\sigma$, which is consistent with the value determined by the first method.

We note that planets too small to be detected are not included in this occurrence calculation.  When including planets smaller than the {\it Kepler}-42 planets within the definition of ``compact multiples,'' this result becomes a lower limit on the occurrence of compact multiples orbiting mid-M dwarfs.

\subsection{Comparison to Sun-like and Early M Dwarf Stars}

We compared this result to the occurrence of compact multiples orbiting Sun-like stars and early M dwarfs.  To create a sample of Sun-like stars, we applied the criteria used by \citet{Petigura2013} to the latest catalog of stellar parameters for the {\it Kepler} targets, compiled by \citet{Huber2014}.  We selected all {\it Kepler} targets with effective temperatures between 4100 and 6100 K, with surface gravities log(g) between 4.0 and 4.9, and with {\it Kepler} magnitudes between 10.0 and 15.0.  We cross-referenced this list of Sun-like {\it Kepler} targets with the NASA Exoplanet Archive list of {\it Kepler} planet candidates and identified 93 systems meeting the criteria for a compact multiple: at least two transiting planet candidates both orbiting with periods of less 10 days.  The systems contain transiting planet candidates with physical radii ranging from 0.45 $R_\Earth$ to 7.0 $R_\Earth$.  With a list of compact multiple systems found in a sample of Sun-like target stars, we repeated the calculation described above.  However, we included a term in the SNR calculation that corrects for the differences in stellar radii between the compact multiple hosts and target stars, using the stellar radii reported in \citet{Huber2014}.  We calculated a compact multiple occurrence of $8.8^{+0.3}_{-0.6}$\% for Sun-like stars, 2.7$\sigma$ lower than the occurrence for mid-M dwarf stars.

As illustrated in Figure \ref{color_color_plot}, early M dwarfs and evolved stars overlap in the $r-J$ vs. $J-K$ color-color diagram.  Instead of using colors to isolate early M dwarf {\it Kepler} targets, instead we used the list of M dwarf targets from \citet[][their Table 4]{Dressing2013}, excluding any M dwarfs with $r-J>3.2$.  We cross-referenced the list to the NASA Exoplanet Archive and found 13 systems with transiting compact multiple candidates: KOI-248, KOI-251, KOI-571, KOI-898, KOI-899, KOI-936, KOI-952, KOI-1078, KOI-1681, KOI-1843, KOI-1867, KOI-2036 and KOI-2793.  As with the Sun-like star calculation, we included a stellar radius correction due to the wide range of radii that early M dwarf stars can have ($\sim$0.3 to 0.6 $R_\Sun)$, taking the stellar radii from \citet{Dressing2013}.  The compact multiple planet candidates orbiting early-M dwarfs range in radii from 0.73 to 2.69 $R_\Earth$, similar to the sizes of compact multiple planets orbiting mid-M dwarfs.  Repeating the calculations described above but for early M dwarfs, we find a compact multiple occurrence rate of $15.9^{+1.5}_{-1.5}$ \%, not significantly different from the occurrence from mid-M dwarf stars.

Figure \ref{fcm} depicts the resulting Monte Carlo distributions of $f_{\rm CM}$ for Sun-like stars, early-M dwarf stars and mid-M dwarf stars.  Although the measured increase in compact multiple occurrence with later spectral type is not statistically significant, it may simply be a consequence of the increase in overall planet occurrence around M dwarf stars first calculated by \citet{Howard2012}.

\section{Discussion}\label{discussion}

Interestingly, {\it Kepler}-445c is somewhat similar in nature to GJ 1214b: both are super-Earths or mini-Neptunes orbiting metal-rich mid-M dwarf stars.  {\it Kepler}-445c orbits with a longer period: 4.87 versus 1.58 days \citep{Carter2011}, and is subject to much lower stellar irradiation. It would be illuminating to study the atmosphere of this planet via transit transmission spectroscopy, as has been done for GJ 1214b \citep{Bean2010b, Croll2011, Desert2011, Berta2012, Fraine2013, Kreidberg2014}, if not for the star's faintness: with a $Ks$ band magnitude of 12.61, transit-transmission spectroscopy will be challenging.

Both the {\it Kepler}-446 and {\it Kepler}-42 systems present cautionary tales for future transit surveys such as NASA's Transit Exoplanet Survey Satellite \citep[TESS,][]{Ricker2014}.  The {\it Kepler} pipeline assigned unphysically large planet-to-star radius ratios with significant uncertainties for the planets in both systems.  In the future, systems like these may have accurate astrometric parallax measurements from ESA's {\it Gaia} Mission.  Combining the {\it Gaia} absolute magnitudes with mass-radius-luminosity relations would enable accurate stellar density priors to be employed during the pipeline transit fitting algorithm.  We suggest such a procedure, as TESS is expected to observe significantly more M dwarf stars than {\it Kepler}.

Following the same calculation in the introduction performed on {\it Kepler}-42, but for {\it Kepler}-445 and {\it Kepler}-446, we calculate that their ``1\% of $M_\star$'' protoplanetary disks consisted of 16.8 and 8.0 $M_\Earth$ of metals, respectively.  In the case of {\it Kepler}-446, where the three planets have radii less than or equal to 1.5 $R_\earth$, we speculate that the three planets have primarily rocky compositions based on the rocky/non-rocky threshold calculated by \citet{Rogers2014}; however, we note that the planet radii are very near the rocky/non-rocky threshold.  Assuming all are rocky, we calculate that the three planets constitute 6.5 $M_\earth$ of metals, meaning that $\sim$80\% of the {\it Kepler}-446's disk metals went into these three planets.  We repeat that this calculation relies on simplistic assumptions, but could indicate that rocky planet formation is efficient around low-mass stars with compact multiple exoplanets.

The same calculation for {\it Kepler}-445 is trickier.  At 2.51 $R_\Earth$, {\it Kepler}-445c is very likely to have a thick gaseous envelope.  However, at 1.58 and 1.25  $R_\Earth$, {\it Kepler}-445b and {\it Kepler}-445d could be rocky, with masses of 3.9 and 1.6 $M_\Earth$, when interpolating onto the \citet{Fortney2007} relations for primarily rocky composition.  That leads to 30\% of the metal content of {\it Kepler}-445's protoplanety disk going into its orbiting planets, not accounting for the core of {\it Kepler}-445c, which is presumably more massive than {\it Kepler}-445b in order to have accreted gas during the protoplanetary disk phase.  Assuming {\it Kepler}-445c has a rocky core as massive as {\it Kepler}-445b, this leads to 56\% of the metal content in these three planets alone.

These results can be interpreted as one or a combination of the following: (1) rocky planet formation is highly efficient around mid-M dwarf stars with compact multiple planetary systems, (2) disk mass fractions around M dwarf stars are significantly higher than 1\% of the host star mass, and/or (3) the planets presented in this work are not in fact rocky, with significant mass in the form of hydrogen and helium.  As stated earlier, next-generation near-infrared precise-radial-velocity spectrometers may be able to confirm the rocky nature of these planets.  If, in fact, rocky planet formation is efficient for compact multiple systems, we would not expect to find significantly more planets beyond the orbits of {\it Kepler}-42 d, {\it Kepler}-445c or {\it Kepler}-446c, simply because there is not enough rocky material in the protoplanetary disk to create significantly more planets.  Therefore, we anticipate a dearth of outer, long-period planets with radii $\geq$ 0.8 $R_\earth$ orbiting mid-M dwarf stars with compact multiple systems.

The planet-metallicity correlation, wherein higher-metallicity stars are more likely to host gaseous planets, has been shown to be true for sun-like stars as well as early M dwarf stars \citep[][]{Santos2001, Fischer2005, Johnson2010, Rojas2010}.  However, it has not been thoroughly investigated for {\it mid}-M dwarf stars due to the challenges in finding planets around them.  It is therefore intriguing that {\it Kepler}-445 \citep[{[Fe/H]}=+0.27$\pm$0.13][]{Muirhead2014} and GJ 1214 \citep[{[Fe/H]}=+0.20][]{Rojas2012} are the only mid-M dwarf stars with transiting planets larger than $2R_\Earth$, and both stars are metal-rich compared to the solar neighborhood \citep[{[Fe/H]}=-0.05][]{Gaidos2014b}.  Based on this alone, we speculate that the planet-metallicity correlation may in fact extend down to mid-M dwarf stars for large planets ($>2R_\Earth$).

\acknowledgements

We thank the anonymous referee for the thoughtful report and useful suggestions.  We thank Andrew West for the suggestions regarding measuring accurate SDSS magnitudes.  A.V. is supported by the National Science Foundation Graduate Research Fellowship under Grant No. DGE1144152. This paper includes data collected by the {\it Kepler} Mission. Funding for the {\it Kepler} Mission is provided by the NASA Science Mission Directorate.  The {\it Kepler} data presented in this paper were obtained from the Mikulski Archive for Space Telescopes (MAST). STScI is operated by the Association of Universities for Research in Astronomy, Inc., under NASA contract NAS5-26555. Support for MAST for non-HST data is provided by the NASA Office of Space Science via grant NNX13AC07G and by other grants and contracts.  

This research has made use of the NASA Exoplanet Archive, which is operated by the California Institute of Technology, under contract with the National Aeronautics and Space Administration under the Exoplanet Exploration Program. This research has made use of the Exoplanet Orbit Database and the Exoplanet Data Explorer at exoplanets.org \citep{Wright2011}.

These results made use of the Discovery Channel Telescope at Lowell Observatory, supported by Discovery Communications, Inc., Boston University, the University of Maryland, the University of Toledo and Northern Arizona University.  The Large Monolithic Imager was funded by the National Science Foundation via grant AST-1005313.  

Some of the data presented herein were obtained at the W.M. Keck Observatory, which is operated as a scientific partnership among the California Institute of Technology, the University of California and the National Aeronautics and Space Administration. The Observatory was made possible by the generous financial support of the W.M. Keck Foundation.  The authors wish to recognize and acknowledge the very significant cultural role and reverence that the summit of Mauna Kea has always had within the indigenous Hawaiian community.

SNIFS on the UH 2.2-m telescope is part of the Nearby Supernova Factory project, a scientific collaboration among the Centre de Recherche Astronomique de Lyon, Institut de Physique NuclŽaire de Lyon, Laboratoire de Physique NuclŽaire et des Hautes Energies, Lawrence Berkeley National Laboratory, Yale University, University of Bonn, Max Planck Institute for Astrophysics, Tsinghua Center for Astrophysics, and the Centre de Physique des Particules de Marseille. 

The infrared spectrum of Barnard's Star was acquired using the SpeX instrument on the Infrared Telescope Facility, which is operated by the University of Hawaii under Cooperative Agreement no. NNX-08AE38A with the National Aeronautics and Space Administration, Science Mission Directorate, Planetary Astronomy Program. 

These results made use of the Sloan Digital Sky Survey (SDSS) III.  Funding for SDSS-III has been provided by the Alfred P. Sloan Foundation, the Participating Institutions, the National Science Foundation, and the U.S. Department of Energy Office of Science. The SDSS-III web site is \url{http://www.sdss3.org/}.  SDSS-III is managed by the Astrophysical Research Consortium for the Participating Institutions of the SDSS-III Collaboration including the University of Arizona, the Brazilian Participation Group, Brookhaven National Laboratory, Carnegie Mellon University, University of Florida, the French Participation Group, the German Participation Group, Harvard University, the Instituto de Astrofisica de Canarias, the Michigan State/Notre Dame/JINA Participation Group, Johns Hopkins University, Lawrence Berkeley National Laboratory, Max Planck Institute for Astrophysics, Max Planck Institute for Extraterrestrial Physics, New Mexico State University, New York University, Ohio State University, Pennsylvania State University, University of Portsmouth, Princeton University, the Spanish Participation Group, University of Tokyo, University of Utah, Vanderbilt University, University of Virginia, University of Washington, and Yale University.  The SDSS data were accessed using the VizieR catalogue access tool, CDS, Strasbourg, France \citep{Ochsenbein2000}.

{\it Facilities:} \facility{DCT (LMI)}, \facility{Hale (TSPEC)}, \facility{IRTF (SpeX)}, \facility{Keck:II (NIRC2, ESI)}, \facility{Kepler}, \facility{UH:2.2m (SNIFS)}

\bibliographystyle{apj}
\bibliography{bibfile}

\end{document}